\documentclass[11pt]{article}
\usepackage{amsmath,amssymb,amsthm}
\usepackage[T1]{fontenc}
\usepackage{lmodern}
\usepackage{amsmath}
\usepackage{cite}
\usepackage{amsthm}
\usepackage[dvips]{graphicx}
\usepackage{hyperref}
\usepackage[usenames, dvipsnames]{color}
\usepackage{mathrsfs}
\usepackage{enumerate}
\usepackage{xcolor}
\definecolor{bl}{rgb}{0.0,0.2,0.6}

\usepackage[]{mdframed}
\usepackage[utf8]{inputenc}

\usepackage{empheq}

\hypersetup{
	unicode=true,           
	pdftoolbar=false,       
	pdfmenubar=false,       
	pdffitwindow=false,     
	pdfstartview={FitH},    
	pdftitle={},            
	pdfauthor={},           
	pdfsubject={},          
	pdfcreator={},          
	pdfproducer={},         
	pdfkeywords={},         
	pdfnewwindow=true,      
	colorlinks=true,        
	linkcolor=bl,           
	citecolor=magenta,      
	filecolor=green,        
	urlcolor=blue           
}

.tfm scaled 1000 
.tfm scaled 1000 
.tfm scaled 1000 

\title{\LARGE\bf Self--dual solutions of a field theory model\\ of two linked rings}

\author{{\sf Neda Abbasi Taklimi}$^{\,a}$\footnote{E-mail: neda.abbasi\_taklimi@phd.usz.edu.pl}\;,
$\;\;\;$ 
{\sf Franco Ferrari}$^{\,a}$\footnote{E-mail: franco@feynman.fiz.univ.szczecin.pl}\;,
$\;\;\;$
{\sf Marcin R.~Pi\c{a}tek}$^{\,a}$\footnote{E-mail: marcin.piatek@usz.edu.pl}
\\[8pt]
$^{a\,}$CASA* and Institute of Physics, University of Szczecin\\ 
Wielkopolska 15, 70--451 Szczecin, Poland
}
\date{}
\begin{document}
	
\maketitle

\begin{abstract}\noindent
In this work the connection established in \cite{F1,FPPZ(2019)NPB} between a model of two 
linked polymers rings with fixed Gaussian linking number forming a 
4-plat  and the statistical mechanics of non-relativistic anyon 
particles is explored.
The excluded volume interactions have been switched off and only the 
interactions of entropic origin arising from the topological constraints 
are considered.
An interpretation from the polymer point of view
of the field equations that minimize the energy of the model
in the limit in which one of the spatial dimensions of the 4-plat 
becomes very large
is provided.
It is shown that the self-dual contributions are responsible for the 
long-range interactions that are necessary for preserving the global 
topological properties of the system during the thermal fluctuations.
The non self-dual part is also related to the topological constraints, and 
takes into account the local interactions acting on the monomers in order 
to prevent the breaking of the polymer lines.
It turns out that the energy landscape of the two linked rings is quite 
complex.
Assuming as a rough approximation that the monomer densities of half of 
the 4-plat are constant, at least two points of energy minimum are 
found.
Classes of non-trivial self-dual solutions of the self-dual field 
equations are derived.
One of these classes is characterized by densities of monomers that are 
the squared modulus of holomorphic functions.
The second class is obtained under some assumptions that allow to reduce 
the self-dual equations to an analog of the Gouy-Chapman equation for the 
charge distribution of ions in a double layer capacitor. In the present 
case, the spatial distribution of the electric potential of the ions is replaced 
by the spatial distribution of the fictitious magnetic fields associated 
with the presence of the topological constraints.
In the limit in which two of the spatial dimensions are large in 
comparison with the third one, we provide exact formulas for the 
conformations of the monomer densities of the 4-plat by using the 
elliptic, hyperbolic and trigonometric solutions of the sinh-Gordon and cosh-Gordon
equations which have been used for instance in the construction of 
classical string solutions in AdS3 and dS3 \cite{Bakas2016}.
\end{abstract}

\newpage
\tableofcontents

\section{Introduction}
In the 70s of the twentieth century, R.G.~de Gennes discovered a relationship between the statistical mechanics of
long polymer molecules and magnetic systems described by multicomponent complex field theories with $O(N)$ symmetry \cite{DeG1,JDC}. 
This discovery gave rise to a new
research direction and resulted in the Nobel Prize in Physics in 1991. 
The idea of using field theory techniques to study properties of polymers was developed by many authors, 
like for instance \cite{VJE,SW,OOF}.
A goal that was set was
to construct a theoretical framework to investigate topologically linked polymers \cite{SFE}. 
This research program has been implemented
in the particular case of polymer rings linked together to form a $2s-$plat
in Refs.~\cite{F1,FPPZ(2019)NPB}.
$2s-$plats are links whose paths in space are characterized by a fixed number  $2s$ of maxima and minima along a given direction called here the height, for instance the $z-$axis.
A physical realization of a $2s-$plat could be a set of polymers rings or knots attached to two membranes consisting of two parallel planes in the $xy$ axis.
It was developed in \cite{FPPZ(2019)NPB} a path integral approach to the statistical mechanics of such a system composed by an arbitrary number of polymer rings linked together. The topological states of the link have been distinguished using a topological invariant known as the
Gauss linking number. The topology of the knots forming the links has been left unspecified. It was shown in \cite{F1,FPPZ(2019)NPB} that the derived path integral formulation
can be mapped into a field theory of quasiparticles known as
anyons. The height $z$ on the polymer side can be interpreted as time in the anyon counterpart. The monomer densities of the $N$ loops become  densities  of a mixture of $N$ types of anyons. When $s=2$, it was demonstrated in \cite{F1} that this field theory admits self-dual solutions.
The integrability of these self-dual configurations have been explored in \cite{FPPZ(2019)NPB}, where a relation with the cosh-Gordon equation has been established.

In this paper, we return to
this subject providing a deeper insight into the meaning of self-duality in the physics of polymers subjected to topological constraints and deriving solutions that minimize the energy of a system of two linked polymer loops.
More in details, using the Bogomol'nyi transformation, it is shown that the energy describing two polymer loops in a 4-plat conformation can be splitted in the limit of long polymers into a self-dual term and a term describing short-range interactions.
Since the interactions have been switched off for simplicity, i.e., the system is in a solution at the theta point, the interactions mentioned here are  of purely topological origin. They are necessary in order to enforce the topological constraints. 
In the limit in which all monomers are equal, i.e., in the case, 
where the 4-plat becomes homopolymeric only the self-dual component survives.
The reason for which a system of homopolymers becomes self-dual is explained in Section \ref{Sect2}.
Next, some exact solutions that minimize the energy of the system at the self-dual point are derived.
We show that the 4-plat has a complex energy landscape. In the case in which the loops are diblock copolymers, we show with a simple approximation (one of the two loops is supposed to have a constant monomer distribution)  that there are at least two points of energy minimum. A class of solutions minimizing the energy in the homopolymeric case consists of field conformations that are holomorphic functions.
Finally, we explore the case in which one of the dimensions, $x$ or $y$, becomes negligible. This can happen for instance in a confined system in which the $y-$dimension is constrained to be small, though it should be big enough to contain the link formed by the two loops, which is intrinsecally three-dimensional.
In this situation, {\it not only} the solutions of the cosh-Gordon equation mentioned above become relevant. 
There are also solutions of the sinh-Gordon and Liouville equations.
In the present paper we consider one-dimensional integrable examples of the first two of these equations.
In this way, starting from certain real ({\it not complex}) solutions of the sinh-/cosh-Gordon equation it has been possible to derive 
exact formulae for (i) field configurations that describe
conformations of the 4-plats; (ii) observables such as monomer densities. 
To achieve this goal the so-called elliptic, hyperbolic and trigonometric solutions are employed. 
These are translationally invariant solutions of the Euclidean sinh-/cosh-Gordon equation 
that depend on {\it only} one variable. These types
of solutions have been used for instance in the construction of classical string solutions in AdS3 and dS3 \cite{Bakas2016}.
Using elliptic, hyperbolic and trigonometric solutions, we find exact formulae for the monomer densities of the 4-plat that minimize the energy at the self-dual point. This is the main result of this work.

This paper is organized as follows.
In Section \ref{Sect2}, we formulate the problem and then, following \cite{F1,FPPZ(2019)NPB}, 
we  re-derive the partition function for a system composed of two linked loops in a topological 
state described by a fixed Gauss linking number.
In the field-theoretical formulation, the polymer partition function can be understood as a correlation function of a mixture of four types of anyons. Referring to the methods coming from the physics of anyons, we derive the self-duality conditions and find their solutions minimizing the energy mentioned above. Then we show that the self-duality equations reduce to the two-dimensional Euclidean sinh-Gordon or cosh-Gordon equations, depending on the sign of the integration constant. We show that there is one more possibility, i.e., when the integration constant equals zero,  we obtain the Liouville equation.\footnote{In the present work, we leave this last case as an open problem for further research.}
In Section \ref{Sect3}, we calculate the translationally invariant solutions of the Euclidean sinh-Gordon and cosh-Gordon equations 
and the polymer densities expressed by them. Here we make an extensive use of the analytical methods developed in \cite{Bakas2016}. 
In Section \ref{Sect4}, we present our conclusions.

\section{Solvable example of topological entanglement}
\label{Sect2}
We consider in this paper links formed in space by two concatenated polymer rings 
with the additional property that the paths of the rings have a fixed number of 
maxima and minima with respect to a particular direction, let's say the direction of the $z$-axis;
$z$ will measure the ``height''. In the case in which the link has a total number of $s$ 
minima and $s$ maxima, the system is called a $2s$-plat. In the following, we will limit ourselves 
to the class of 4-plats which is particularly interesting for biological applications \cite{DWS}. 
Since $s=2$, each ring has only one point of maximum and only one point of minimum.
\begin{figure}[t]
	\centering
	\includegraphics[scale=0.49]{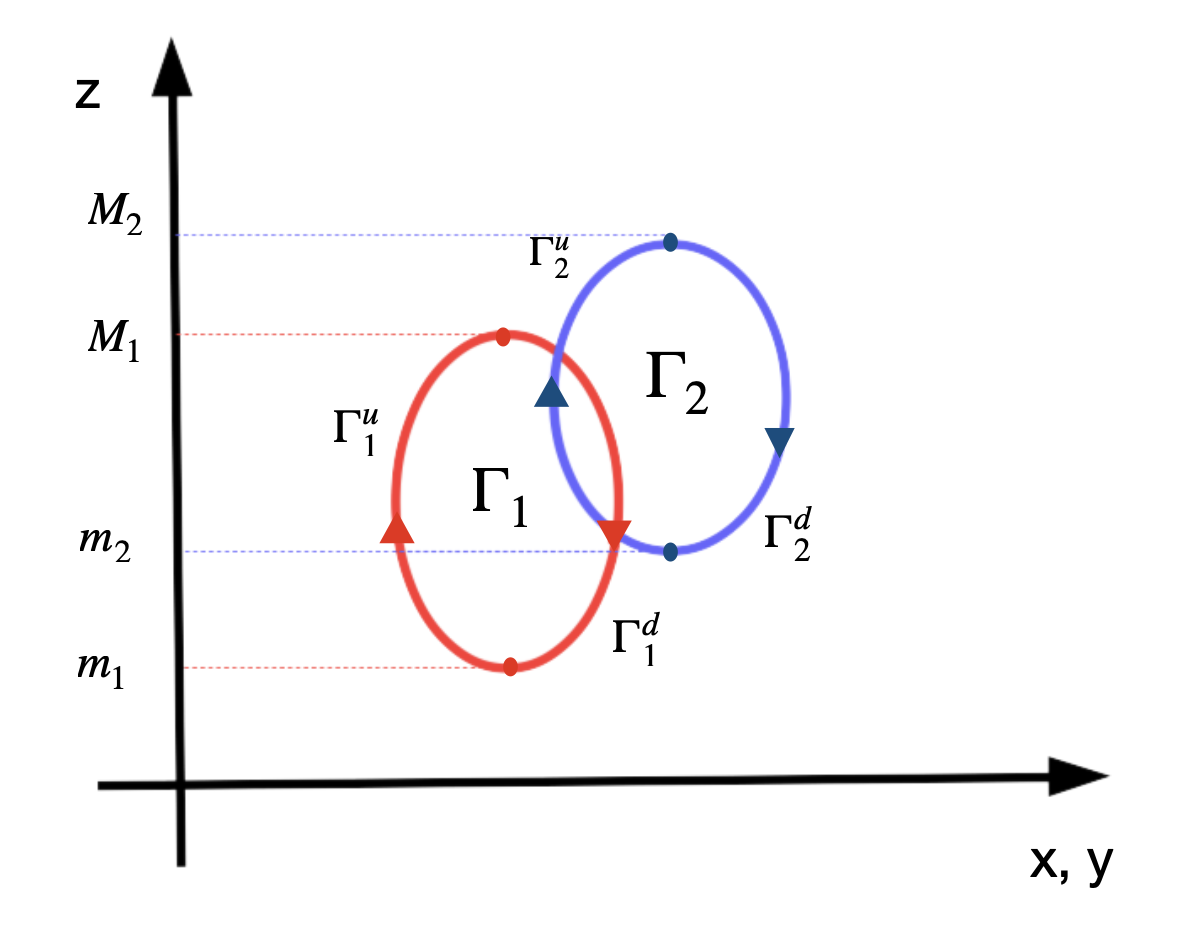}
	\caption{\it The 4-plat in our parametrization.}
	\label{fig4plat}
\end{figure}
Let $\Gamma_1$ denote the path of the first ring and $\Gamma_2$ the path of the second one. 
Here, $m_a$ and $M_a$, $a=1,2$ will be respectively the points of minimum and maximum of $\Gamma_a$ (see Fig.\ref{fig4plat}).
Each loop $\Gamma_a$ will be further decomposed into two monotonic curves $\Gamma_{a}^{u}$ and $\Gamma_{a}^{d}$.
The loops $\Gamma_{a}^{u}$ run upwards in the $z$-direction and $\Gamma_{a}^{d}$ run downwards according to 
the orientation of the loops given in Fig.\ref{fig4plat}.

In order to distinguish the different topologies of the link the Gauss linking number $\chi(\Gamma_1,\Gamma_2)$ 
will be used, though the treatment can be generalized to the more powerful Vassiliev invariants.
As shown in Ref.~\cite{Hornig}, in the case of the decomposition of two loops into monotonic curves 
$\Gamma_{a}^{u}$, $\Gamma_{a}^{d}$, $a=1,2$
along a preferred  direction $z$, $\chi(\Gamma_1,\Gamma_2)$ can be written as follows\footnote{See also \cite{FPPZ(2019)NPB}.}
\begin{equation}\label{GLNW}
\chi(\Gamma_1,\Gamma_2)\;=\;
W_{\Gamma_{1}^{u}\Gamma_{2}^{d}}(z_0,z_1)+
W_{\Gamma_{1}^{u}\Gamma_{2}^{u}}(z_0,z_1)+
W_{\Gamma_{1}^{d}\Gamma_{2}^{u}}(z_0,z_1)+
W_{\Gamma_{1}^{d}\Gamma_{2}^{d}}(z_0,z_1),
\end{equation}
where $W_{\Gamma\Gamma'}(z_0,z_1)$ is the winding number of two monotonic curves $\Gamma$, $\Gamma'$
between the two heights $z_0$ and $z_1$:
\begin{equation}\label{windef}
W_{\Gamma\Gamma'}(z_0,z_1)\;=\;\epsilon_{ij}\int\limits_{z_0}^{z_1}{\rm d}\!\left(x^{i}(z)-x^{\prime i}(z)\right)
\frac{\left(x^{j}(z)-x^{\prime j}(z)\right)}{\left|{\boldsymbol x}(z)-{\boldsymbol x}^{\prime}(z)\right|^2}.
\end{equation}
Since $z$ is a privileged direction, we distinguish the two-dimensional spatial components of the coordinates 
$x^{i}$, $i=1,2$ and the $z$ component $x^{0}$.
In general, vectors will be denoted with the symbol $({\boldsymbol v},v^{0})$, where ${\boldsymbol v}=(v^{1},v^{2})$
is the projection of the vector in the $xy$-plane.
Note that the two curves $\Gamma$, $\Gamma'$ in Eq.~(\ref{windef}) do not need to be defined in the same 
interval of heights. For this reason, the integration limits in the definition (\ref{windef}) of the winding
number are in the interval $[z_0,z_1]$ in which both curves have points at the same height.
For instance, in the situation of Fig.\ref{fig4plat} in which $m_1<m_2<M_1<M_2$ we have that
$z_0=m_2$ and $z_1=M_1$.

The partition function $Z(\mu)$ of the system composed by the two linked loops $\Gamma_1$, $\Gamma_2$
may be written as follows,
\begin{equation}\label{partbas}\boxed{
Z(\mu)\;=\;\left[\prod\limits_{a=1}^{2}
\int\limits_{{\boldsymbol x}_{a}(m_a)}^{{\boldsymbol x}_{a}(M_a)}
{\cal D}{\boldsymbol x}_{a}^{u}(z)
\int\limits_{{\boldsymbol x}_{a}(M_a)}^{{\boldsymbol x}_{a}(m_a)}
{\cal D}{\boldsymbol x}_{a}^{d}(z)
\right]\delta\left(\chi(\Gamma_1,\Gamma_2)-\mu\right)
{\rm e}^{-A_{\sf pol}},}
\end{equation}
the coordinates ${\boldsymbol x}_{a}(m_a)$ and ${\boldsymbol x}_{a}(M_a)$ 
denoting respectively the locations of the points of maximal and minimal height 
of $\Gamma_{a}$ are fixed. Moreover, the topological constraint 
\begin{equation}
\chi(\Gamma_1,\Gamma_2)\;=\;\mu
\end{equation}
with $\mu$ being a constant is imposed in Eq.~(\ref{partbas}) using a Dirac delta function.
Finally, $A_{\sf pol}$ is the term associated to chain connectivity:
\begin{equation}
A_{\sf pol}\;=\;\prod\limits_{a=1}^{2}\int\limits_{m_a}^{M_a}{\rm d}z
\left[g_{a,u}\left|\frac{{\rm d}{\boldsymbol x}_{a}^{u}(z)}{{\rm d}z}\right|^2
+g_{a,d}\left|\frac{{\rm d}{\boldsymbol x}_{a}^{d}(z)}{{\rm d}z}\right|^2\right].
\end{equation}
For simplicity, no interactions have been added, though the treatment can be
easily extended to include the excluded volume potential.
In the above formulae ${\boldsymbol x}_{a}^{u,d}(z)$ are curves describing the paths of $\Gamma_{a}^{u,d}$.
The quantities $g_{a,u}$'s and $g_{a,d}$'s are constants related to the Kuhn length and 
characterizing the flexibility of the chains $\Gamma_{a}^{u}$ and $\Gamma_{a}^{d}$.

At this point, the Fourier transform can be applied to represent the Dirac delta function 
$\delta\left(\chi(\Gamma_1,\Gamma_2)-\mu\right)$ in the following form:
\begin{equation}
\delta\left(\chi(\Gamma_1,\Gamma_2)-\mu\right)\;=\; 
\int\limits_{-\infty}^{+\infty}{\rm d}\lambda\;{\rm e}^{{\rm i}\lambda\mu}
\;{\rm e}^{-{\rm i}\lambda\chi(\Gamma_1,\Gamma_2)}.
\end{equation}
This allows to rewrite the partition function $Z(\mu)$ in the simpler form:
\begin{equation}\boxed{
Z(\mu)\;=\; 
\int\limits_{-\infty}^{+\infty}{\rm d}\lambda\;{\rm e}^{{\rm i}\lambda\mu}\;Z(\lambda),}
\end{equation}
where
\begin{equation}\boxed{
Z(\lambda)\;=\; \left[\prod\limits_{a=1}^{2}
\int\limits_{{\boldsymbol x}_{a}(m_a)}^{{\boldsymbol x}_{a}(M_a)}
{\cal D}{\boldsymbol x}_{a}^{u}(z)
\int\limits_{{\boldsymbol x}_{a}(M_a)}^{{\boldsymbol x}_{a}(m_a)}
{\cal D}{\boldsymbol x}_{a}^{d}(z)
\right]\,{\rm e}^{-A_{\sf pol}}\,{\rm e}^{-{\rm i}\lambda\chi(\Gamma_1,\Gamma_2)}.}
\end{equation}
The Fourier transformation from $Z(\mu)$ to $Z(\lambda)$ is an anologue of the passage
from the microcanonical ensemble to the canonical ensemble, but the role of the Hamiltonian 
$H$ is replaced here by the Gauss linking number $\chi(\Gamma_1,\Gamma_2)$ and 
the Boltzmann factor $\beta=(kT)^{-1}$ is replaced by ${\rm i}\lambda$.

As shown in Ref.~\cite{FPPZ(2019)NPB}, the exponential ${\rm e}^{-{\rm i}\lambda\chi(\Gamma_1,\Gamma_2)}$,
which contains a very complicated dependence on the conformations ${\boldsymbol x}_{a}^{u,d}(z)$,
may be simplified rewriting it as the partition function of an abelian BF-model:
\begin{eqnarray}\label{reltop}
{\rm e}^{-{\rm i}\lambda\chi(\Gamma_1,\Gamma_2)}&=&
\int{\cal D}{\boldsymbol B}_{1}({\boldsymbol x},t)
{\cal D}{\boldsymbol B}_{2}({\boldsymbol x},t)
{\cal D}B_{1}^{0}({\boldsymbol x},t)
{\cal D}B_{2}^{0}({\boldsymbol x},t)\nonumber\\
&&\hspace{-50pt}\;\times 
\exp\left\lbrace-{\rm i}S_{\sf BF}-{\rm i}\lambda\int{\rm d}^2x{\rm d}t\left[{\boldsymbol B}_{2}\cdot{\boldsymbol J}_{1} 
+B_{2}^{0}J_{1}^{0}\right]-\frac{{\rm i}\kappa}{8\pi^2}
\int{\rm d}^2x{\rm d}t\left[{\boldsymbol B}_{1}\cdot{\boldsymbol J}_{2} 
+B_{1}^{0}J_{2}^{0}\right]\right\rbrace,
\end{eqnarray}
where
\begin{equation}
S_{\sf BF}\;=\;\frac{\kappa}{4\pi}\epsilon_{ij}\int{\rm d}^2x{\rm d}t
\left[B^{0}_{1}\partial^{i}B^{j}_{2}+B^{0}_{2}\partial^{i}B^{j}_{1}\right],
\quad\quad i,j=1,2.
\end{equation}
Here and below it is assumed that repeated upper and lower indices that label spatial coordinates are summed.
The sources of the magnetic field $({\boldsymbol B}_{a},B^{0}_{a})$, $a=1,2$ 
appearing in Eq.~(\ref{reltop}) are imaginary currents flowing inside the loops $\Gamma_1$ and $\Gamma_2$:
\begin{eqnarray}
{\boldsymbol J}_{a}({\boldsymbol x},t)&=&\int\limits_{m_a}^{M_a}{\rm d}z
\frac{{\rm d}{\boldsymbol x}_{a}^{u}(z)}{{\rm d}z}\delta^{(2)}\!\left({\boldsymbol x}-{\boldsymbol x}_{a}^{u}(z)\right)
\delta(t-z)
+\int\limits_{m_a}^{M_a}{\rm d}z\;
\frac{{\rm d}{\boldsymbol x}_{a}^{d}(z)}{{\rm d}z}\delta^{(2)}\!\left({\boldsymbol x}-{\boldsymbol x}_{a}^{d}(z)\right)
\delta(t-z),\\
J_{a}^{0}({\boldsymbol x},t)&=&\int\limits_{m_a}^{M_a}{\rm d}z\,
\delta^{(2)}\!\left({\boldsymbol x}-{\boldsymbol x}_{a}^{u}(z)\right)\delta(t-z)
+\int\limits_{m_a}^{M_a}{\rm d}z\,
\delta^{(2)}\!\left({\boldsymbol x}-{\boldsymbol x}_{a}^{d}(z)\right)\delta(t-z).
\end{eqnarray}
Let's notice that in Eq.~(\ref{reltop}) the fields ${\boldsymbol B}_{a}({\boldsymbol x},t)$
satisfy the Coulomb gauge condition:
\begin{equation}\label{cgauge}
{\boldsymbol\nabla}\cdot{\boldsymbol B}_{a}({\boldsymbol x},t)\;=\;0.
\end{equation}
The Coulomb gauge arises naturally when the curves $\Gamma_1$, $\Gamma_2$ are parametrized using
the $z$ coordinate and divided into monotonic curves $\Gamma_{a}^{u,d}$, $a=1,2$.
Indeed, it is easy to show that 
\begin{equation}
{\boldsymbol\nabla}\cdot{\boldsymbol J}_{a}({\boldsymbol x},t)\;=\;0,
\end{equation}
so that the longitudinal component of the currents in the $xy$-plane is vanishing.
Thus, the longitudinal component of the magnetic fields ${\boldsymbol B}_{a}({\boldsymbol x},t)$
have no sources and the condition (\ref{cgauge}) is automatically satisfied.

In order to prove Eq.~(\ref{reltop}), one needs to integrate out the fields 
${\boldsymbol B}_{a}({\boldsymbol x},t)$, $B_{a}^{0}({\boldsymbol x},t)$
on the right hand side of that equation. This amounts to a Gaussian integration that 
may be easily performed using the propagator:
\begin{equation}
\left\langle B_{1}^{0}({\boldsymbol x},t)B_{2}^{i}({\boldsymbol y},t')\right\rangle
\;=\;-\left\langle B_{1}^{i}({\boldsymbol x},t)B_{2}^{0}({\boldsymbol y},t')\right\rangle
\;=\;\frac{\delta(t-t')}{2\kappa}\,\epsilon_{ij}\partial_{y^j}\ln\left|{\boldsymbol x}-{\boldsymbol y}\right|^2.
\end{equation}

Applying the identity (\ref{reltop}) it is possible to convert the partition function $Z(\lambda)$ 
to the following form:
\begin{equation}\label{zupdown}\boxed{
Z(\lambda)\;=\;\int\prod\limits_{a=1}^{2}
{\cal D}{\boldsymbol B}_{a}{\cal D}B_{a}^{0}\,
Z_{a}^{u}(\lambda)\,Z_{a}^{d}(\lambda)\,{\rm e}^{-{\rm i}S_{\sf BF}},}
\end{equation}	
where
\begin{eqnarray}
Z_{a}^{u}(\lambda)&=&\int\limits_{{\boldsymbol x}_{a}(m_a)}^{{\boldsymbol x}_{a}(M_a)}
{\cal D}{\boldsymbol x}_{a}^{u}(z)\,{\rm e}^{-S_{a}^{u}},\\
Z_{a}^{d}(\lambda)&=&\int\limits_{{\boldsymbol x}_{a}(M_a)}^{{\boldsymbol x}_{a}(m_a)}
{\cal D}{\boldsymbol x}_{a}^{d}(z)\,{\rm e}^{-S_{a}^{d}}
\end{eqnarray}
and
\begin{eqnarray}
S_{a}^{u}&=&\int\limits_{m_a}^{M_a}{\rm d}z
\left[g_{a,u}\left|\frac{{\rm d}{\boldsymbol x}_{a}^{u}(z)}{{\rm d}z}\right|^2
+{\rm i}\sum_{b=1}^{2}C_{ab}
\left(\frac{{\rm d}{\boldsymbol x}_{a}^{u}(z)}{{\rm d}z}\cdot
{\boldsymbol B}_{b}({\boldsymbol x}_{a}^{u}(z),z)+B_{b}^{0}({\boldsymbol x}_{a}^{u}(z),z)\right)\right],\\
S_{a}^{d}&=&\int\limits_{m_a}^{M_a}{\rm d}z
\left[g_{a,d}\left|\frac{{\rm d}{\boldsymbol x}_{a}^{d}(z)}{{\rm d}z}\right|^2
-{\rm i}\sum_{b=1}^{2}C_{ab}
\left(\frac{{\rm d}{\boldsymbol x}_{a}^{d}(z)}{{\rm d}z}\cdot
{\boldsymbol B}_{b}({\boldsymbol x}_{a}^{d}(z),z)+B_{b}^{0}({\boldsymbol x}_{a}^{d}(z),z)\right)\right].
\end{eqnarray}	
The $2\times 2$ matrix $C_{ab}$ is given by
$$
C_{ab}\;=\;\begin{bmatrix} 
	\;0 & \lambda\; \\
	\;\frac{\kappa}{8\pi^2} & 0\; \\
\end{bmatrix}.
$$
Let us note $S_{a}^{u}$ and $S_{a}^{d}$ are formally equal to the actions of two particles
immersed in the magnetic fields generated by the vector potentials 
${\boldsymbol B}_{1}$, ${\boldsymbol B}_{2}$ and interacting with the external potentials
$B_{1}^{0}$, $B_{2}^{0}$. Accordingly, $Z_{a}^{u}(\lambda)$ may be interpreted as the transition
amplitudes of particles ${\boldsymbol x}_{a}^{u}(z)$ to pass from an initial state $|\,{\boldsymbol x}_{a}^{u}(m_a)\,\rangle$
to a final state $\langle\,{\boldsymbol x}_{a}^{u}(M_a)\,|$. An analogous interpretation can be given to $Z_{a}^{d}(\lambda)$.
This analogy with quantum mechanics allows to pass from paths to fields using the procedure of second quantisation.

Putting
\begin{equation}
Z_{a}^{u,d}(\lambda)\;=\;G_{a}^{u,d}\!\left({\boldsymbol x}_{a}(M_a)-{\boldsymbol x}_{a}(m_a),M_a-m_a\right)
\end{equation}
it is possible to show that the one-particle transition amplitudes $Z_{a}^{u,d}(\lambda)$
satisfy the pseudo-Schr\"{o}dinger equations:
\begin{eqnarray}\label{pschrodu}
&&\left[\frac{\partial}{\partial t}-{\rm i}\sum_{b=1}^{2}C_{ab}B_{b}^{0}({\boldsymbol x},t)
-\frac{1}{4g_{a,u}}\left({\boldsymbol\nabla}_{{\boldsymbol x}}-{\rm i}\sum_{b=1}^{2}C_{ab}
{\boldsymbol B}_{b}({\boldsymbol x},t)\right)^2\right]
G_{a}^{u}\!\left({\boldsymbol x}-{\boldsymbol y},t-t'\right)\nonumber\\
&&\hspace{250pt}\;=\;\delta^{(2)}\!\left({\boldsymbol x}-{\boldsymbol y}\right)\delta\left(t-t'\right),\\
&&\left[\frac{\partial}{\partial t}+{\rm i}\sum_{b=1}^{2}C_{ab}B_{b}^{0}({\boldsymbol x},t)
-\frac{1}{4g_{a,d}}\left({\boldsymbol\nabla}_{{\boldsymbol x}}+{\rm i}\sum_{b=1}^{2}C_{ab}
{\boldsymbol B}_{b}({\boldsymbol x},t)\right)^2\right]
G_{a}^{d}\!\left({\boldsymbol x}-{\boldsymbol y},t-t'\right)\nonumber\\
&&\hspace{250pt}\;=\;\delta^{(2)}\!\left({\boldsymbol x}-{\boldsymbol y}\right)\delta\left(t-t'\right).
\label{pschrodd}
\end{eqnarray}
It turns out from Eqs.~(\ref{pschrodu}) and (\ref{pschrodd}) that $Z_{a}^{u,d}(\lambda)$
are the Green functions of a set of complex scalar fields $\psi_{a}^{u,d}$, $\psi_{a}^{*\,u,d}$, i.e.,
\begin{eqnarray}\label{Zu}
Z_{a}^{u}(\lambda)&=&\frac{1}{Z_{a}^{u}}\int{\cal D}\psi_{a}^{*\,u}({\boldsymbol x},t){\cal D}\psi_{a}^{u}({\boldsymbol x},t)
\,\psi_{a}^{*\,u}\!\left({\boldsymbol x}(M_a),M_a\right)\psi_{a}^{u}\left({\boldsymbol x}(m_a),m_a\right)\nonumber\\
&&\;\hspace{-50pt}\times 
\exp\left\lbrace-\int{\rm d}^2x{\rm d}t\left[\psi_{a}^{*\,u}
\left(\partial^{0}-{\rm i}\sum_{b=1}^{2}C_{ab}B_{b}^{0}\right)\psi_{a}^{u}
+\frac{1}{4g_{a,u}}\left|\left({\boldsymbol\nabla}_{{\boldsymbol x}}-{\rm i}\sum_{b=1}^{2}C_{ab}
{\boldsymbol B}_{b}\right)\psi_{a}^{u}\right|^2\right]\right\rbrace
\end{eqnarray}
and
\begin{eqnarray}\label{Zd}
	Z_{a}^{d}(\lambda)&=&\frac{1}{Z_{a}^{d}}\int{\cal D}\psi_{a}^{*\,d}({\boldsymbol x},t)
	{\cal D}\psi_{a}^{d}({\boldsymbol x},t)
	\,\psi_{a}^{*\,d}\!\left({\boldsymbol x}(m_a),m_a\right)\psi_{a}^{d}\left({\boldsymbol x}(M_a),M_a\right)\nonumber\\
&&\;\hspace{-50pt}\times
	\exp\left\lbrace-\int{\rm d}^2x{\rm d}t\left[\psi_{a}^{*\,d}
	\left(\partial^{0}+{\rm i}\sum_{b=1}^{2}C_{ab}B_{b}^{0}\right)\psi_{a}^{d}
	+\frac{1}{4g_{a,d}}\left|\left({\boldsymbol\nabla}_{{\boldsymbol x}}+{\rm i}\sum_{b=1}^{2}C_{ab}
	{\boldsymbol B}_{b}\right)\psi_{a}^{d}\right|^2\right]\right\rbrace
\end{eqnarray}
with $Z_{a}^{u,d}$ being the partition functions of the complex scalar fields,
\begin{eqnarray}\label{Zaufields}
Z_{a}^{u}&=&
\int{\cal D}\psi_{a}^{*\,u}({\boldsymbol x},t){\cal D}\psi_{a}^{u}({\boldsymbol x},t)
\nonumber\\
&&\;\hspace{-50pt}\times
\exp\left\lbrace-\int{\rm d}^2x{\rm d}t\left[\psi_{a}^{*\,u}
\left(\partial^{0}-{\rm i}\sum_{b=1}^{2}C_{ab}B_{b}^{0}\right)\psi_{a}^{u}
+\frac{1}{4g_{a,u}}\left|\left({\boldsymbol\nabla}_{{\boldsymbol x}}-{\rm i}\sum_{b=1}^{2}C_{ab}
{\boldsymbol B}_{b}\right)\psi_{a}^{u}\right|^2\right]\right\rbrace
\end{eqnarray}
and
\begin{eqnarray}\label{Zadfields}
Z_{a}^{d}&=&
\int{\cal D}\psi_{a}^{*\,d}({\boldsymbol x},t){\cal D}\psi_{a}^{d}({\boldsymbol x},t)
\nonumber\\
&&\;\hspace{-50pt}\times
\exp\left\lbrace-\int{\rm d}^2x{\rm d}t\left[\psi_{a}^{*\,d}
\left(\partial^{0}+{\rm i}\sum_{b=1}^{2}C_{ab}B_{b}^{0}\right)\psi_{a}^{d}
+\frac{1}{4g_{a,d}}\left|\left({\boldsymbol\nabla}_{{\boldsymbol x}}+{\rm i}\sum_{b=1}^{2}C_{ab}
{\boldsymbol B}_{b}\right)\psi_{a}^{d}\right|^2\right]\right\rbrace.
\end{eqnarray}
The integrations over $t$ are made over the intervals $\left[m_a,M_a\right]\ni t$.

Further processing of the expression (\ref{zupdown}) into a more useful form is a bit difficult 
because of the factors $(Z_{a}^{u})^{-1}$ and $(Z_{a}^{d})^{-1}$ in (\ref{Zu}) and (\ref{Zd}). 
However, it can be dealt with using the so-called replica method.
Indeed, by introducing $n$ replica fields:
\begin{equation}
{\boldsymbol\Psi}_{a}^{u,d}\;=\;
\left(\psi_{a}^{u,d(1)},\ldots,\psi_{a}^{u,d(n)}\right),
\quad\quad
{\boldsymbol\Psi}_{a}^{*\,u,d}\;=\;
\left(\psi_{a}^{*\,u,d(1)},\ldots,\psi_{a}^{*\,u,d(n)}\right),
\end{equation}
the partition function in Eq.~(\ref{zupdown}) may be rewritten as a product of Gaussian integrals,
\begin{empheq}[box=\fbox]{align}\label{zlrep}
	Z(\lambda)&\;=\;\lim\limits_{n\to 0}\int\prod\limits_{a=1}^2
	{\cal D}{\boldsymbol B}_{a}{\cal D}B_{a}^{0}
	{\cal D}{\boldsymbol\Psi}_{a}^{*\,u}
	{\cal D}{\boldsymbol\Psi}_{a}^{u}
	{\cal D}{\boldsymbol\Psi}_{a}^{*\,d}
	{\cal D}{\boldsymbol\Psi}_{a}^{d}
	\nonumber\\[3pt]
	&\;\times\;
	\psi_{a}^{*\,u(1)}\!\left({\boldsymbol x}(M_a),M_a\right)
	\psi_{a}^{u(1)}\!\left({\boldsymbol x}(m_a),m_a\right)
	\psi_{a}^{*\,d(1)}\!\left({\boldsymbol x}(m_a),m_a\right)
	\psi_{a}^{d(1)}\!\left({\boldsymbol x}(M_a),M_a\right)
	\nonumber\\[3pt]
	&\;\times\;
	\exp\left\lbrace-\int{\rm d}^2x{\rm d}t\left[{\boldsymbol\Psi}_{a}^{*\,u}
	\left(\partial^{0}-{\rm i}\sum_{b=1}^{2}C_{ab}B_{b}^{0}\right){\boldsymbol\Psi}_{a}^{u}
	+\frac{1}{4g_{a,u}}\left|\left({\boldsymbol\nabla}_{{\boldsymbol x}}-{\rm i}\sum_{b=1}^{2}C_{ab}
	{\boldsymbol B}_{b}\right){\boldsymbol\Psi}_{a}^{u}\right|^2\right]\right\rbrace
	\nonumber\\[3pt]
	&\;\times\; 
	\exp\left\lbrace-\int{\rm d}^2x{\rm d}t\left[{\boldsymbol\Psi}_{a}^{*\,d}
	\left(\partial^{0}+{\rm i}\sum_{b=1}^{2}C_{ab}B_{b}^{0}\right){\boldsymbol\Psi}_{a}^{d}
	+\frac{1}{4g_{a,d}}\left|\left({\boldsymbol\nabla}_{{\boldsymbol x}}+{\rm i}\sum_{b=1}^{2}C_{ab}
	{\boldsymbol B}_{b}\right){\boldsymbol\Psi}_{a}^{d}\right|^2\right]\right\rbrace
	\nonumber\\[3pt]
	&\;\times\;
	{\rm e}^{-{\rm i}S_{\sf BF}}.
\end{empheq}
In this form the $z$-components $B_{a}^{0}$ of the vector potentials play the role of 
Lagrange multipliers. They may be easily integrated out producing the constraints:\footnote{The following notation  
	$\left|{\boldsymbol\Psi}_{a}^{u,d}\right|^2=\sum_{r=1}^{n}\psi_{a}^{u,d(r)}\psi_{a}^{*\,u,d(r)}$ is 
	employed in Eqs.~(\ref{CC1}) and (\ref{CC2}).}
\begin{eqnarray}
	\frac{\kappa}{4\pi}\epsilon_{ij}\partial^i B^{j}_{2}&=&
C_{21}\left(
	-\left|{\boldsymbol\Psi}_{2}^{u}\right|^2+\left|{\boldsymbol\Psi}_{2}^{d}\right|^2
	\right)
	\theta(M_2-t)\theta(t-m_2),
	\label{CC1}\\[3pt]
	\frac{\kappa}{4\pi}\epsilon_{ij}\partial^i B^{j}_{1}&=&
C_{12}\left(
-\left|{\boldsymbol\Psi}_{1}^{u}\right|^2+\left|{\boldsymbol\Psi}_{1}^{d}\right|^2
\right)
\theta(M_1-t)\theta(t-m_1)\label{CC2}
\end{eqnarray}
or
\begin{equation}\label{gconstr}
\sum\limits_{c=1}^{2}d_{ac}\epsilon^{ij}\partial_{i}B_{c,j}=
\sum\limits_{b=1}^{2}C_{ba}\left(-\left|{\boldsymbol\Psi}_{b}^{u}\right|^2+\left|{\boldsymbol\Psi}_{b}^{d}\right|^2\right)
\theta(M_b-t)\theta(t-m_b),
\end{equation}
where
$$
d_{ab}\;=\;\begin{bmatrix} 
	\;0 & \frac{\kappa}{4\pi} \\
	\;\frac{\kappa}{4\pi} & 0\; \\
\end{bmatrix}.
$$

The Heaviside theta functions are necessary in order to take into account the fact
that the loops $\Gamma_{1}$ and $\Gamma_{2}$ are defined in the different ranges of heights 
$\left[m_1,M_1\right]$ and $\left[m_2,M_2\right]$.
Putting into (\ref{CC1}) and (\ref{CC2}) the Gauss law constraint, i.e.,
\begin{equation}
B^{j}_{a}\;=\;\epsilon^{jk}\partial_{k}\phi_a
\end{equation}
one gets relations typical for electrostatics:
\begin{eqnarray}\label{es1}
\Delta\phi_2 &=&\frac{1}{2n}\left(
-\left|{\boldsymbol\Psi}_{2}^{u}\right|^2+\left|{\boldsymbol\Psi}_{2}^{d}\right|^2
\right)
\theta(M_2-t)\theta(t-m_2),
\\[3pt]
\Delta\phi_1 &=&\frac{4n\lambda}{\kappa}
\left(
-\left|{\boldsymbol\Psi}_{1}^{u}\right|^2+\left|{\boldsymbol\Psi}_{1}^{d}\right|^2
\right)
\theta(M_1-t)\theta(t-m_1).\label{es2}
\end{eqnarray}
Let us notice that $\lambda\in\left[-\infty,\infty\right]$ and the change of sign of $\lambda$
reverses the sign of the density of charges $\left|{\boldsymbol\Psi}_{1}^{u}\right|^2$ and
$\left|{\boldsymbol\Psi}_{1}^{d}\right|^2$ in Eq.~(\ref{es2}).
The partition function $Z(\lambda)$ in Eq.~(\ref{zlrep}) originates from a polymer problem,
but may also be interpreted as the correlation function of a mixture of four types
of anyon particles with densities 
$\left|{\boldsymbol\Psi}_{a}^{u}\right|^2$ and
$\left|{\boldsymbol\Psi}_{a}^{d}\right|^2$, $a=1,2$ and the action:
\begin{eqnarray}\label{Sanyon}
S_{\sf matter}&=&\sum\limits_{a=1}^{2}\int{\rm d}^2x\int\limits_{0}^{T}{\rm d}t
\left[
{\boldsymbol\Psi}_{a}^{*\,u}\partial_{0}{\boldsymbol\Psi}_{a}^{u}
+\frac{1}{4g_{a,u}}\left|\left({\boldsymbol\nabla}_{{\boldsymbol x}}-{\rm i}\sum_{b=1}^{2}C_{ab}
{\boldsymbol B}_{b}\right){\boldsymbol\Psi}_{a}^{u}\right|^2
\right.\nonumber\\
&&\hspace{66pt}+\left.
{\boldsymbol\Psi}_{a}^{*\,d}\partial_{0}{\boldsymbol\Psi}_{a}^{d}
+\frac{1}{4g_{a,d}}\left|\left({\boldsymbol\nabla}_{{\boldsymbol x}}+{\rm i}\sum_{b=1}^{2}C_{ab}
{\boldsymbol B}_{b}\right){\boldsymbol\Psi}_{a}^{d}\right|^2
\right].
\end{eqnarray}
For simplicity, from now on we assume that $m_1=m_2=0$ and $M_1=M_2=T$.
This implies that the Heaviside theta functions in the constraints (\ref{CC1}) and (\ref{CC2}) 
satisfied by the vector potentials ${\boldsymbol B}_{a}$ are no longer necessary.

The analogy with anyons suggests to investigate the action (\ref{Sanyon}) with 
the methods of self-dual systems. Following \cite{Dunne1,Dunne2} and \cite{F1}, we introduce
to this purpose the covariant derivatives:
\begin{eqnarray}
D_{a,j}^{u}&=&\partial_{j}-{\rm i}\sum_{b=1}^{2}C_{ab}{\boldsymbol B}_{b,j},\\
D_{a,j}^{d}&=&\partial_{j}+{\rm i}\sum_{b=1}^{2}C_{ab}{\boldsymbol B}_{b,j},
\end{eqnarray}
where $j=1,2$ labels the spatial coordinates and $a=1,2$ labels the contributions coming from loops 
$\Gamma_1$ and $\Gamma_2$. One gets the following identities:
\begin{eqnarray}\label{Du}
\left|{\boldsymbol D}_{a}^{u}{\boldsymbol\Psi}_{a}^{u}\right|^2 &=&
\left|\left(D_{a,1}^{u}\pm {\rm i}D_{a,2}^{u}\right){\boldsymbol\Psi}_{a}^{u}\right|^2
\mp\left|{\boldsymbol\Psi}_{a}^{u}\right|^2\sum_{b=1}^{2}C_{ab}
\left(\partial_2B_{b,1}-\partial_1B_{b,2}\right)\mp\epsilon^{ij}\partial_{i}J_{a,j}^{u},
\\
\left|{\boldsymbol D}_{a}^{d}{\boldsymbol\Psi}_{a}^{d}\right|^2 &=&
\left|\left(D_{a,1}^{d}\pm {\rm i}D_{a,2}^{d}\right){\boldsymbol\Psi}_{a}^{d}\right|^2
\mp\left|{\boldsymbol\Psi}_{a}^{d}\right|^2\sum_{b=1}^{2}C_{ab}
\left(\partial_2B_{b,1}-\partial_1B_{b,2}\right)\mp\epsilon^{ij}\partial_{i}J_{a,j}^{d},
\label{Dd}
\end{eqnarray}
where
\begin{eqnarray}
J_{a,j}^{u}&=&{\boldsymbol\Psi}_{a}^{*\,u}D_{a,j}^{u}{\boldsymbol\Psi}_{a}^{u},
\\
J_{a,j}^{d}&=&{\boldsymbol\Psi}_{a}^{*\,d}D_{a,j}^{d}{\boldsymbol\Psi}_{a}^{d}.
\end{eqnarray}
Note that the magnetic fields $\partial_2B_{b,1}-\partial_1B_{b,2}$ along the $z$-axis
can be expressed in (\ref{Du}) and (\ref{Dd}) by the densities as a result of use of the constraints (\ref{CC1}) and (\ref{CC2}).
Putting all together, it is possible to rewrite the action (\ref{Sanyon}) as follows
\begin{equation}\label{action3}
S_{\sf matter}\;=\;I_{\sf T}+I_{\sf sd}+I_{\sf C}\;,
\end{equation}
where
\begin{equation}\label{action31}
I_{\sf T}\;=\;\sum\limits_{a=1}^{2}\int{\rm d}^2x\int\limits_{0}^{T}{\rm d}t
\left[ 
{\boldsymbol\Psi}_{a}^{*\,u}\partial_{0}{\boldsymbol\Psi}_{a}^{u}+
{\boldsymbol\Psi}_{a}^{*\,d}\partial_{0}{\boldsymbol\Psi}_{a}^{d}
\right],
\end{equation}
$I_{\sf sd}$ is the self-dual part of the action,
\begin{equation}\label{action32}
I_{\sf sd}\;=\;\sum\limits_{a=1}^{2}\int{\rm d}^2x\int\limits_{0}^{T}{\rm d}t
\left[ 
\frac{1}{4g_{a,u}}\left|\left(D_{a,1}^{u}+{\rm i}D_{a,2}^{u}\right){\boldsymbol\Psi}_{a}^{u}\right|^2
+\frac{1}{4g_{a,d}}\left|\left(D_{a,1}^{d}+{\rm i}D_{a,2}^{d}\right){\boldsymbol\Psi}_{a}^{d}\right|^2
\right]
\end{equation}	
and $I_{\sf C}$ accounts for the Coulomb-like interactions,
\begin{eqnarray}\label{action33}
I_{\sf C}&=&\frac{\lambda}{8\pi}\int{\rm d}^2x\int\limits_{0}^{T}{\rm d}t
\left[\left(-\frac{1}{g_{1,u}}\left|{\boldsymbol\Psi}_{1}^{u}\right|^2
+\frac{1}{g_{1,d}}\left|{\boldsymbol\Psi}_{1}^{d}\right|^2\right)
\left(-\left|{\boldsymbol\Psi}_{2}^{u}\right|^2+\left|{\boldsymbol\Psi}_{2}^{d}\right|^2\right)
\right.\nonumber\\
&&\hspace{85pt}+\left.
\left(\frac{1}{g_{2,u}}\left|{\boldsymbol\Psi}_{2}^{u}\right|^2
-\frac{1}{g_{2,d}}\left|{\boldsymbol\Psi}_{2}^{d}\right|^2\right)
\left(-\left|{\boldsymbol\Psi}_{1}^{u}\right|^2+\left|{\boldsymbol\Psi}_{1}^{d}\right|^2\right)\right].
\end{eqnarray}	
In writing the action in Eqs.~(\ref{action3})-(\ref{action33}) terms that are total derivatives have been neglected.

It turns out that the term $I_{\sf T}$ is negligibly small when the height $\tau$ in which 
the paths of both loops are defined become large.
To show this, it is sufficient to perform in Eqs.~(\ref{action31}), (\ref{action32}) and (\ref{action33}) 
the change of variable $\tau\sigma=t$. After doing this the self-dual contribution $I_{\sf sd}$ 
and the Coulomb interaction terms $I_{\sf C}$ pick up a factor $\tau$, but not $I_{\sf T}$. 
From now on we will working in the limit of large $\tau$, in  which
\begin{equation}\label{action4}
	S_{\sf matter}\;\sim\;I_{\sf sd}+I_{\sf C}\;.
\end{equation}
Apart from the fact that the variable $t$ describes the height of the polymer loops and not time, 
the above action is formally equal to that of a systems of anyons.
In other words, as it is reasonable to expect, in the limit $\tau\to\infty$ the monomer 
distribution does not change very much at different heights, so that it becomes possible 
to talk about static solutions similarly to the case of anyons with the difference that here static meanas absence of
changes along the $z$-axis.

Proceeding analogously as in the case of anyons, on the basis of Eq.~(\ref{action4}) we define the density of 
energy per unit of height $z$,
\begin{eqnarray}\label{energy1}
{\cal E}(z) &=& 
\sum\limits_{a=1}^{2}\int{\rm d}^2x
\left[ 
\frac{1}{4g_{a,u}}\left|\left(D_{a,1}^{u}+{\rm i}D_{a,2}^{u}\right){\boldsymbol\Psi}_{a}^{u}\right|^2
+\frac{1}{4g_{a,d}}\left|\left(D_{a,1}^{d}+{\rm i}D_{a,2}^{d}\right){\boldsymbol\Psi}_{a}^{d}\right|^2
\right]\nonumber\\
&+&
\frac{\lambda}{8\pi}\int{\rm d}^2x
\left[\left(-\frac{1}{g_{1,u}}\left|{\boldsymbol\Psi}_{1}^{u}\right|^2
+\frac{1}{g_{1,d}}\left|{\boldsymbol\Psi}_{1}^{d}\right|^2\right)
\left(-\left|{\boldsymbol\Psi}_{2}^{u}\right|^2+\left|{\boldsymbol\Psi}_{2}^{d}\right|^2\right)
\right.\nonumber\\
&&\hspace{85pt}+\left.
\left(\frac{1}{g_{2,u}}\left|{\boldsymbol\Psi}_{2}^{u}\right|^2
-\frac{1}{g_{2,d}}\left|{\boldsymbol\Psi}_{2}^{d}\right|^2\right)
\left(-\left|{\boldsymbol\Psi}_{1}^{u}\right|^2+\left|{\boldsymbol\Psi}_{1}^{d}\right|^2\right)\right].
\end{eqnarray}
An interesting case is when the monomer densities of $\Gamma_{a}^{u}$ or $\Gamma_{a}^{u}$ 
can be considered as constant. For instance, assuming that $|{\boldsymbol\Psi}_{1}^{d}|^2=V_{1}^{2}$ 
and $|{\boldsymbol\Psi}_{2}^{d}|^2=V_{2}^{2}$  with $V_1,V_2={\rm const.}$, we obtain
\begin{eqnarray}\label{energy2}
	{\cal E}(z) &=& 
	{\cal E}_{\sf sd}(z)\nonumber\\
	&+&
	\frac{\lambda}{8\pi}\int{\rm d}^2x
	\left[\left(\frac{1}{g_{1,u}}\left|{\boldsymbol\Psi}_{1}^{u}\right|^2
	-\frac{1}{g_{1,d}}V_{1}^{2}\right)
	\left(\left|{\boldsymbol\Psi}_{2}^{u}\right|^2-V_{2}^{2}\right)
	\right.\nonumber\\
	&&\hspace{85pt}+\left.
	\left(\frac{1}{g_{2,u}}\left|{\boldsymbol\Psi}_{2}^{u}\right|^2
	-\frac{1}{g_{2,d}}V_{2}^{2}\right)
	\left(-\left|{\boldsymbol\Psi}_{1}^{u}\right|^2+V_{1}^{2}\right)\right],
\end{eqnarray}
where ${\cal E}_{\sf sd}(z)$ is the energy density of the self-dual part,
\begin{equation}\label{energysd}
{\cal E}_{\sf sd}(z)\;=\;\sum\limits_{a=1}^{2}\int{\rm d}^2x
\left[\frac{1}{4g_{a,u}}\left|\left(D_{a,1}^{u}+{\rm i}D_{a,2}^{u}\right){\boldsymbol\Psi}_{a}^{u}\right|^2
+\frac{1}{4g_{a,d}}\left|\left(D_{a,1}^{d}+{\rm i}D_{a,2}^{d}\right){\boldsymbol\Psi}_{a}^{d}\right|^2
\right].
\end{equation}
The energy in Eq.~(\ref{energy2}) is minimized by the self-duality conditions:
\begin{equation}\label{sdcond}\boxed{
\left(\,D_{a,1}^{u,d}+{\rm i}D_{a,2}^{u,d}\,\right){\boldsymbol\Psi}_{a}^{u,d}\;=\;0}
\end{equation}
which are satisfied. There are two distinct minima corresponding to the following cases:
\begin{equation}
\left|{\boldsymbol\Psi}_{1}^{u}\right|^2\;=\;\frac{g_{1,u}}{g_{1,d}}V_{1}^{2}
\quad {\rm and} \quad
\left|{\boldsymbol\Psi}_{2}^{u}\right|^2\;=\;\frac{g_{2,u}}{g_{2,d}}V_{2}^{2}
\end{equation}
or
\begin{equation}
	\left|{\boldsymbol\Psi}_{1}^{u}\right|^2\;=\;V_{1}^{2}
	\quad {\rm and} \quad
	\left|{\boldsymbol\Psi}_{2}^{u}\right|^2\;=\;V_{2}^{2}.
\end{equation}
Most interesting is probably the homopolymer case in which 
all legs $\Gamma_{1}^{u,d}$ and $\Gamma_{2}^{u,d}$ are homogeneous, so that
\begin{equation}\label{homcond}
\frac{1}{g_{1,u}}\;=\;\frac{1}{g_{1,d}}\;=\;\frac{1}{g_{2,u}}\;=\;\frac{1}{g_{2,d}}\;=\;\frac{1}{g}.
\end{equation}
Remarkably, if all parameters $g_{a,u}$ and $g_{a,d}$ are equal, then the Coulomb-like short range
interactions disappear and the system becomes self-dual, i.e., ${\cal E}(z)={\cal E}_{\sf sd}(z)$.
The vanishing of the short-range interactions reminds the case of solutions at high monomer concentration and good
solvents, in which the interactions act on each monomer symmetrically from any direction, so that their total
effect is negligible. 
The situation is similar here. 
The term $I_{\sf C}$ of Eq.~(\ref{action33}) accounts for the short range interactions 
and they vanish in the limit $\tau\to\infty$ and when the loops are homogeneous, 
see condition (\ref{homcond}). As already mentioned, in the large $\tau$ limit the monomer
distribution is not depending on the $z$ direction implying that the short range forces due to 
the topological constraints acting on a monomer from above are counterbalanced by the forces acting from below.
In the $xy$ directions all legs $\Gamma_{a}^{u}$ and $\Gamma_{a}^{d}$ are equal under the conditions (\ref{homcond}).
It is thus likely that the short range interactions of topological origin become isotropic as in the case of
polymer solutions at high monomer concentrations. Of course, what is not cancelled are the long-range 
interactions because they are necessary to keep the topology of the link.
These long term interactions are taken into account by the self-dual contributions in Eq.~(\ref{action32}).

In the remaining part of this work some solutions of the self-duality equations (\ref{sdcond}) will be derived.
First, we consider the self-duality equations for the fields ${\boldsymbol\Psi}_{a}^{u}$, i.e., 
\begin{equation}\label{saux}
\left(\partial_{1}-{\rm i}\sum_{b=1}^{2}C_{ab}{\boldsymbol B}_{b,1}\right){\boldsymbol\Psi}_{a}^{u}
+{\rm i}\left(\partial_{2}-{\rm i}\sum_{b=1}^{2}C_{ab}{\boldsymbol B}_{b,2}\right){\boldsymbol\Psi}_{a}^{u}
\;=\;0.
\end{equation}
We attempt the ansatz
\begin{equation}
{\boldsymbol\Psi}_{a}^{u}\;=\;{\boldsymbol\Psi}_{a}^{u}(x^1+{\rm i}x^2),
\end{equation} 
i.e., ${\boldsymbol\Psi}_{a}^{u}$ is a holomorphic function of the complex variable $w=x^1+{\rm i}x^2$.
With this setting it turns out that 
$\partial_{1}{\boldsymbol\Psi}_{a}^{u}+{\rm i}\partial_{2}{\boldsymbol\Psi}_{a}^{u}=0$.
In this way Eq.~(\ref{saux}) simplifies to
\begin{equation}\label{cond2}
-{\rm i}\sum_{b=1}^{2}C_{ab}{\boldsymbol B}_{b,1}{\boldsymbol\Psi}_{a}^{u}+
\sum_{b=1}^{2}C_{ab}{\boldsymbol B}_{b,2}{\boldsymbol\Psi}_{a}^{u}\;=\;0.
\end{equation}
The fields may be derived by solving the Gauss constraints (\ref{gconstr}),
\begin{equation}
B_{c,j}\;=\;\epsilon_{jk}\sum\limits_{a=1}^{2}\left(d^{-1}\right)_{ca}\sum\limits_{b=1}^{2}C_{ba}
\int{\rm d}^2y\ln\left|{\boldsymbol x}-{\boldsymbol y}\right|
\left(|{\boldsymbol\Psi}_{b}^{d}|^2-|{\boldsymbol\Psi}_{b}^{u}|^2\right).
\end{equation}
Clearly, Eq.~(\ref{cond2}) is satisfied if $|{\boldsymbol\Psi}_{b}^{u}|^2=|{\boldsymbol\Psi}_{b}^{d}|^2$.
The derivation of the expression of the fields ${\boldsymbol\Psi}_{b}^{d}$ is straightforward.

A connection between the polymer problem treated here and nonlinear models will be established in the following.
Since we are now considering the energy ${\cal E}(z)$ coming from the partition function $Z_{\sf PF}$ associated
to the Green function (\ref{zlrep}),
\begin{equation}
\label{Zpf}
Z_{\sf PF}\;=\;\int\prod\limits_{a=1}^{2}
{\cal D}{\boldsymbol B}_{a}{\cal D}B_{a}^{0}\,
Z_{a}^{u}\,Z_{a}^{d}\,{\rm e}^{-{\rm i}S_{\sf BF}},
\end{equation}
the appropriate expression of the energy is obtained in the limit $n\to 1$ in which the number of replicas is equal to one.
In Eq.~(\ref{Zpf}) the subpartition functions $Z_{a}^{u}$, $Z_{a}^{d}$ are given by Eqs.~(\ref{Zaufields}) and (\ref{Zadfields}).
Putting all together in the field equations (\ref{sdcond}) and the Gauss constraints (\ref{gconstr}), 
we obtain the equations for $\psi_{a}^{u,d}$, $\psi_{a}^{*\,u,d}$, ${\boldsymbol B}_{a,i}$, i.e.:
\begin{eqnarray}\label{eq1}
&&\left(\partial_1-{\rm i}\sum\limits_{b=1}^{2}C_{ab}{\boldsymbol B}_{b,1}\right)\psi_{a}^{u}	
+ {\rm i}\left(\partial_2-{\rm i}\sum\limits_{b=1}^{2}C_{ab}{\boldsymbol B}_{b,2}\right)\psi_{a}^{u}\;=\; 0,
\\
\label{eq2}
&&\left(\partial_1+{\rm i}\sum\limits_{b=1}^{2}C_{ab}{\boldsymbol B}_{b,1}\right)\psi_{a}^{d}	
+ {\rm i}\left(\partial_2-{\rm i}\sum\limits_{b=1}^{2}C_{ab}{\boldsymbol B}_{b,2}\right)\psi_{a}^{d}\;=\; 0
\end{eqnarray}
and
\begin{equation}
\label{eq3}
\sum\limits_{c=1}^{2}d_{ac}\epsilon^{ij}\partial_{i}{\boldsymbol B}_{c,j}
\;=\;\sum\limits_{b=1}^{2}C_{ba}\left(-|{\boldsymbol\Psi}_{b}^{u}|^2
+|{\boldsymbol\Psi}_{b}^{d}|^2\right).
\end{equation}
At this point we perform in Eqs.~(\ref{eq1})-(\ref{eq3}) the transformation:
\begin{equation}\label{trsf1}
\psi_{a}^{u,d}\;=\;\sqrt{\rho_{a}^{u,d}}\;{\rm e}^{{\rm i}\theta_{a}^{u,d}}.
\end{equation}
From (\ref{eq1})-(\ref{eq3}) we obtain $4+1$ equations after separating the real and
imaginary terms:
\begin{eqnarray}\label{a1}
\frac{1}{2}\partial_{1}\log\rho_{a}^{u}-\partial_{2}\theta_{a}^{u}+
\sum\limits_{b=1}^{2}C_{ab}{\boldsymbol B}_{b,2} &=& 0,
\\
\label{a2}
\partial_{1}\theta_{a}^{u}-\sum\limits_{b=1}^{2}C_{ab}{\boldsymbol B}_{b,1}+
\frac{1}{2}\partial_{2}\log\rho_{a}^{u} &=& 0,
\\
\label{a3}
\frac{1}{2}\partial_{1}\log\rho_{a}^{d}-\partial_{2}\theta_{a}^{d}-
\sum\limits_{b=1}^{2}C_{ab}{\boldsymbol B}_{b,2} &=& 0,
\\
\label{a4}
\partial_{1}\theta_{a}^{d}+\sum\limits_{b=1}^{2}C_{ab}{\boldsymbol B}_{b,1}+
\frac{1}{2}\partial_{2}\log\rho_{a}^{d} &=& 0,
\end{eqnarray}
\begin{equation}\label{a5}
\sum\limits_{c=1}^{2}d_{ac}\epsilon^{ij}\partial_{i}{\boldsymbol B}_{c,j}\;=\;
\sum\limits_{b=1}^{2}C_{ba}\left(-\rho_{b}^{u}+\rho_{b}^{d}\right).
\end{equation}
By requiring that the expressions of ${\boldsymbol B}_{b,1}$ calculated using Eqs.~(\ref{a2}) and (\ref{a4})
are the same we obtain the consistency conditions:
\begin{equation}
\partial_{1}\theta_{a}^{u}+\frac{1}{2}\partial_{2}\log\rho_{a}^{u}\;=\;
-\partial_{1}\theta_{a}^{d}-\frac{1}{2}\partial_{2}\log\rho_{a}^{d}.
\end{equation}
These conditions are satisfied if we require that
\begin{equation}\label{b1}\boxed{
\theta_{a}^{u}\;=\;-\theta_{a}^{d}
\quad{\rm and}\quad
\rho_{a}^{u}\;=\;\frac{A_{a}}{\rho_{a}^{d}}.}
\end{equation}
The $A_a$ are real constants that may be positive or negative.
The same conditions may be found by checking the consistency of Eqs.~(\ref{a1}) and (\ref{a3}).
Thanks to Eqs.~(\ref{b1}) we may concentrate to the calculation of $\theta_{a}^{u}$, $\rho_{a}^{u}$
and ${\boldsymbol B}_{a,1}$, ${\boldsymbol B}_{a,2}$. 
The latter quantities are known by solving Eqs.~(\ref{a1}),  (\ref{a2}) and (\ref{a5}).
In order to eliminate the vector potentials ${\boldsymbol B}_{a,i}$ we differentiate (\ref{a1}) with respect to $x^1$ and (\ref{a2})
with respect to $x^2$:
\begin{eqnarray}
\sum\limits_{b=1}^{2}C_{ab}\partial_1{\boldsymbol B}_{b,2}&=&\partial_1\partial_2\theta_{a}^{u}
-\frac{1}{2}\partial_{1}^{2}\log\rho_{a}^{u},\\
\sum\limits_{b=1}^{2}C_{ab}\partial_2{\boldsymbol B}_{b,1}&=&\partial_1\partial_2\theta_{a}^{u}
+\frac{1}{2}\partial_{2}^{2}\log\rho_{a}^{u}.
\end{eqnarray}
Subtracting the first of the above equations from the second one, we get
\begin{equation}\label{c1}
\sum\limits_{b=1}^{2}C_{ab}\left(\partial_1{\boldsymbol B}_{b,2}-\partial_2{\boldsymbol B}_{b,1}\right)
\;=\;-\frac{1}{2}\left(\Delta\log\rho_{a}^{u}\right),
\end{equation}
where $\Delta=\partial_{1}^{2}+\partial_{2}^{2}$ is the Laplacian.
Remembering that $\epsilon^{ij}\partial_{i}{\boldsymbol B}_{b,j}=\partial_1{\boldsymbol B}_{b,2}-\partial_2{\boldsymbol B}_{b,1}$
and substituting (\ref{c1}) in Eq.~(\ref{a5}) we obtain the final result:
\begin{equation}
-\frac{1}{2}\sum\limits_{c,a=1}^{2}\left(C^{-1}\right)_{ca}\left(\Delta\log\rho_{a}^{u}\right)d_{ec}
\;=\;\sum\limits_{b=1}^{2}C_{be}\left(-\rho_{b}^{u}+\frac{A_b}{\rho_{b}^{u}}\right),
\end{equation}
where $\left(C^{-1}\right)_{ab}$ is the inverse of the matrix $C_{ab}$, i.e., 
$$
\left(C^{-1}\right)_{ab}\;=\;\begin{bmatrix} 
	\;0 & \frac{8\pi^2}{\kappa}\; \\
	\;\lambda^{-1} & 0\; \\
\end{bmatrix}.
$$
Making explicit the dependence on the physical parameters, one gets:
\begin{eqnarray}
\Delta\log\rho_{1}^{u}&=&\frac{\lambda}{\pi}\left(\rho_{2}^{u}-\frac{A_2}{\rho_{2}^{u}}\right),\\
\Delta\log\rho_{2}^{u}&=&\frac{\lambda}{\pi}\left(\rho_{1}^{u}-\frac{A_1}{\rho_{1}^{u}}\right).
\end{eqnarray}
The constants $A_1$, $A_2$ may take any real value.
Due to the fact that the two loops $\Gamma_1$ and $\Gamma_2$ have the same properties and lengths by assumption,
it is natural to search for solutions such that 
\begin{equation}\label{Asamp}\boxed{
\rho_{1}^{u}\;=\;\rho_{2}^{u}\quad {\rm and}\quad A_1\;=\;A_2.}
\end{equation}
In this case, what remains to be solved is the following equation for $\rho_{1}^{u}$, i.e.,
\begin{equation}\label{r1u}\boxed{
\Delta\log\rho_{1}^{u}\;=\;\frac{\lambda}{\pi}\left(\rho_{1}^{u}-\frac{A_1}{\rho_{1}^{u}}\right).}
\end{equation}
Taking $A_1>0$ in (\ref{r1u}), i.e., $A_1=|A_1|$ and putting:
\begin{equation}\label{eta}\boxed{
\eta\;=\;\log\left(\frac{\rho_{1}^{u}}{\sqrt{|A_1|}}\right)\;\;\Leftrightarrow\;\;
\rho_{1}^{u}\;=\;\sqrt{|A_1|}{\rm e}^{\eta}.}
\end{equation}
one gets
\begin{equation}\boxed{
\Delta\eta\;=\;\frac{\lambda\sqrt{|A_1|}}{\pi}\left({\rm e}^{\eta}-{\rm e}^{-\eta}\right).}
\label{SGE}
\end{equation}
This is the equation of sinh-Gordon.
Taking $A_1<0$ in (\ref{r1u}), i.e., $A_1=-|A_1|$ and putting (\ref{eta}) one gets
\begin{equation}
\boxed{\Delta\eta\;=\;\frac{\lambda\sqrt{|A_1|}}{\pi}\left({\rm e}^{\eta}+{\rm e}^{-\eta}\right).}
\label{CGE}
\end{equation}
This is the cosh-Gordon equation.
Thus, Eq.~(\ref{r1u}) can be reduced to the sinh-Gordon and
cosh-Gordon equations, respectively, depending on the sign of the parameter $A_1$.
There is one more possibility, i.e., taking $A_1=0$ in Eq.~(\ref{r1u}) and substituting $\rho_{1}^{u}={\rm e}^{\phi}$.
This leads to the Liouville equation,  
$\Delta\phi=\frac{\lambda}{\pi}{\rm e}^{\phi}$.
In this situation, however, the calculation of the density $\rho_{1}^{u}$
is more sophisticated (some singularities appear) but rather feasible ({\it cf}.~\cite{JP}).
We leave this case as an open problem for further research.\footnote{
These are new observations that have not been spelled out in \cite{FPPZ(2019)NPB}.}

\section{Translationally invariant solutions}
\label{Sect3}

\subsection{Elliptic solutions of sinh-Gordon and cosh-Gordon equations}
In this subsection we construct solutions to Eqs.~(\ref{SGE}) and (\ref{CGE}), 
which depend on only one variable, {\it cf}.~\cite{Bakas2016}.\footnote{
In the present section we take advantage of  calculations presented in \cite{Bakas2016}.}
To begin with, let us write Eqs.~(\ref{SGE}) and (\ref{CGE}) in the unified form
\begin{equation}
\partial_{1}^{2}\eta+\partial_{2}^{2}\eta
\;=\;\frac{m^2}{2}\left({\rm e}^{\eta}+t{\rm e}^{-\eta}\right),\;\;{\rm where}\;\;m^2=\pi^{-1}2\lambda\sqrt{|A_1|}
\label{OurCGeq2}
\end{equation}
and $t=+1$ for the cosh-Gordon equation, and $t=-1$ for the sinh-Gordon equation.
Here we are searching for solutions of Eq.~(\ref{OurCGeq2}) in the 
form\footnote{Note, coordinates here are with subscripts. The theory is Euclidean, so it doesn't matter.}
\begin{equation}\label{e1}
\eta(x_1,x_2)\;=\;\eta_{1}(x_1),
\end{equation}
i.e., the solutions
translationally invariant in the $x_2$ direction. (For translationally invariant solutions 
in the $x_1$ direction, there is exactly the same analysis and results as will be discussed below.)
In this case, Eq.~(\ref{OurCGeq2}) reduces to the ordinary differential equation (ODE),
\begin{equation}
	\frac{{\rm d}^2\eta_1}{{\rm d}x_{1}^{2}}
	\;=\;\frac{m^2}{2}\left({\rm e}^{\eta_1}+t{\rm e}^{-\eta_1}\right),
	\label{OurCGode}
\end{equation}
which can be integrated to 
\begin{equation}\label{IntCGode}
\frac{1}{2}\left(\frac{{\rm d}\eta_1}{{\rm d}x_{1}}\right)^2-\frac{m^2}{2}\left({\rm e}^{\eta_1}
-t{\rm e}^{-\eta_1}\right)\;=\;E.
\end{equation}
The quantity $E$ in Eq.~(\ref{IntCGode}) is an arbitrary constant.

Substituting, 
\begin{equation}\label{v1alfa}
\;v_1\;=\;\frac{m^2}{2}{\rm e}^{\eta_1},
\end{equation}
in (\ref{IntCGode}) one gets
\begin{equation}\label{IntV1}
\left(\frac{{\rm d}v_1}{{\rm d}x_{1}}\right)^2\;=\;2v_{1}^{3}+2Ev_{1}^{2}-t\frac{m^4}{2}v_1.
\end{equation}
Finally, after one more substitution, 
\begin{equation}\label{V1}
\;v_1\;=\;2y-\frac{1}{3}E,
\end{equation}
Eq.~(\ref{IntV1}) takes the standard Weierstrass form:
\begin{eqnarray}\label{Wei}
\left(\frac{{\rm d}y}{{\rm d}x_{1}}\right)^2 &=& 
4y^3-\left(\frac{1}{3}E^2+t\frac{m^4}{4}\right)y
+\frac{E}{3}\left(\frac{1}{9}E^2+t\frac{m^4}{8}\right)\nonumber\\
&\equiv& 4y^3-g_2y-g_3
\end{eqnarray}
with constants $g_2$, $g_3$ fixed as
\begin{eqnarray}\label{g2g3}
g_2(E,t)&=&\frac{1}{3}E^2+t\frac{m^4}{4},\nonumber\\
g_3(E,t)&=&-\frac{E}{3}\left(\frac{1}{9}E^2+t\frac{m^4}{8}\right).
\end{eqnarray}	
Our goal is to find real solutions of the Weierstrass Eq.~(\ref{Wei})
defined in the real domain. Indeed, it is noticeable that Eq.~(\ref{OurCGeq2}) 
is real. We recall that $\eta$ was introduced in (\ref{eta}) as a logarithm of the positive function $\rho_{11}$.
Hence, $\eta$ and then $y$ in Eq.~(\ref{Wei}) must be real-valued.

Before we proceed to the construction of real solutions of Eq.~(\ref{Wei}), 
let us recall some properties of the Weierstrass elliptic function $\wp(z;g_2,g_3)$.\footnote{We refer here to the work \cite{Bakas2016}.}
The latter is a doubly periodic complex function of one complex variable $z$,
satisfying the complex domain Weierstrass ODE,
\begin{equation}\label{Wp}
\left(\frac{{\rm d}y}{{\rm d}z}\right)^2=4y^3-g_2y-g_3.
\end{equation}
The periods of $\wp(z;g_2,g_3)$ are related to the three roots $e_1$, $e_2$, $e_3$  
of the cubic polynomial $Q(y)=4y^3-g_2y-g_3$.
Let us introduce the discriminant $\Delta\equiv g_{2}^{3}-27g_{3}^{2}$.
One can distinguish three cases.
\begin{enumerate}
\item
If $\Delta>0$ then the cubic polynomial $Q(y)$ will have three real roots.
For $e_1>e_2>e_3$ the function $\wp$ has one real period $2\omega_1$ and one
imaginary period $2\omega_2$ which are related to the roots:
\begin{equation}
\omega_1=\frac{K(k)}{\sqrt{e_1-e_3}}, 
\quad
\omega_2=\frac{{\rm i}K(k')}{\sqrt{e_1-e_3}}.
\end{equation}
Here, $K(k)$ is the complete elliptic integral of the first kind and
\begin{equation}
k^2=\frac{e_2-e_3}{e_1-e_3}, 
\quad
k'^{2}=\frac{e_1-e_2}{e_1-e_3},
\quad
k^2+k'^{2}=1.
\end{equation}
\item
If $\Delta<0$ then the cubic polynomial $Q(y)$ will have one real root and two complex,
which are conjugate to each other. Let $e_2$ be the real root and $e_{1,3}=a\pm{\rm i}b$
with $b>0$. Then, the function $\wp$ has one real period and one complex, which is not purely imaginary.
In this case it is convenient to consider as fundamental periods the complex one and its conjugate:
\begin{equation}
\omega_1=\frac{K(k)-{\rm i}K(k')}{2\sqrt[4]{9a^2+b^2}},
\quad
\omega_2=\frac{K(k)+{\rm i}K(k')}{2\sqrt[4]{9a^2+b^2}}
\end{equation}
with
\begin{equation}
k^2=\frac{1}{2}-\frac{3e_2}{4\sqrt{9a^2+b^2}},
\quad
k'^2=\frac{1}{2}+\frac{3e_2}{4\sqrt{9a^2+b^2}}.
\end{equation}
The real period is just the sum of the two fundamental periods $2\omega_1$ and $2\omega_2$.
\item
If $\Delta=0$ then at least two of the roots are equal and the Weierstrass function $\wp$ takes
a special form, which is not doubly periodic, but trigonometric.
\end{enumerate}
\begin{figure}[t]
	\centering
	\includegraphics[scale=0.5]{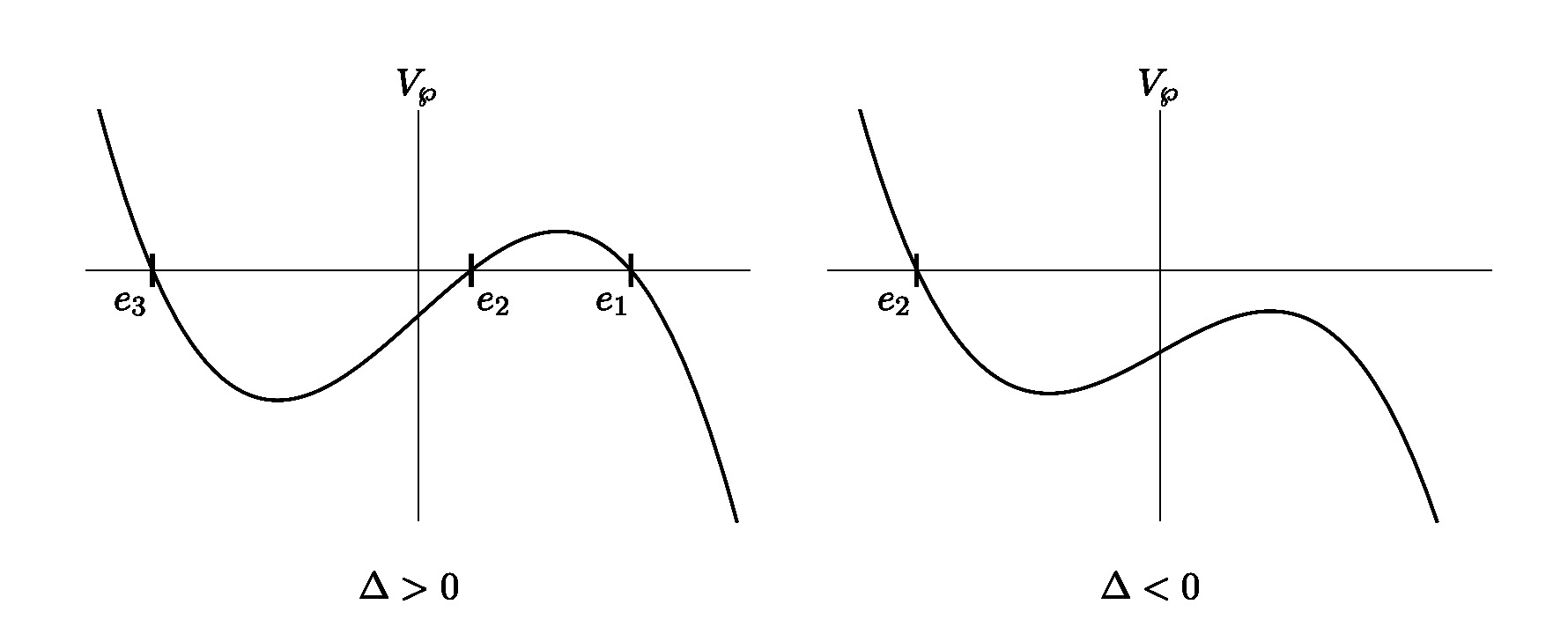}
	\caption{\it The cubic potential $V_{\wp}(y)$ for $\Delta>0$ and $\Delta<0$ ({\it cf}.~\cite{Bakas2016}).}
	\label{fig1}
\end{figure}

\noindent
In cases 1. and 2. the Weierstrass function $\wp$ obeys the following half-period relations
$\wp(\omega_1)=e_1$, $\wp(\omega_2)=e_3$, $\wp(\omega_3)=e_2$, where $\omega_3\equiv\omega_1+\omega_2$.

It is known how to derive real solutions of Eq.~(\ref{Wp}).
As spotted in \cite{Bakas2016}, Eq.~(\ref{Wp}) can be understood as the conservation of energy for the 
one-dimensional classical mechanical system, namely, for a point particle with vanishing energy moving 
under the influence of the cubic potential $V_{\wp}(y)=-4y^3+g_{2}y+g_{3}=-Q(y)$.
\begin{itemize}
\item[$\circ$] If $\Delta>0$ there exist two real solutions of Eq.~(\ref{Wp}), one being unbounded with $y>e_1$ 
and a bounded one with $e_3<y<e_2$ (see Fig.\ref{fig1}).
\item[$\circ$] If $\Delta<0$ there exists only 
one real solution of Eq.~(\ref{Wp}), which is unbounded with $y>e_2$.
\item[$\circ$] The unbounded real solutions are given by $\wp(x)$, where $x\in\mathbb{R}$.
Indeed, the Weierstrass function $\wp$ has a second order pole at $z=0$ and it is real on the real axis.
So, to determine the unbounded real solution it is enough to restrict $z$ to the real axis.
\item[$\circ$] The bounded real solutions are determined by $\wp(x+\omega_2)$, where $x\in\mathbb{R}$.
Precisely, $\wp(x+\omega_2)$ is a real periodic solution that oscillates between $e_2$ and $e_3$, 
with period equal to $2\omega_1$. The proof of this statement is as follows ({\it cf}.~\cite{Bakas2016}).
First, it is clear that the argument of the Weierstrass function can be shifted by an 
arbitrary constant and $\wp$ still solves Eq.~(\ref{Wp}). Second, using the periodic properties of 
$\wp$ and the fact that $\omega_2$ is purely imaginary in the case 
$\Delta>0$, one gets the relation:
\begin{equation}
	\overline{\wp(x+\omega_2)}=\wp(x+\overline{\omega_2})=\wp(x-\omega_2)=\wp(x+\omega_2).
\end{equation}
Finally, the above claim stems from the fact that the Weierstrass function obeys 
the half-period identities  $\wp(\omega_1+\omega_2)=e_2$ and
$\wp(\omega_2)=e_3$.
\end{itemize}
To conclude, when $\Delta>0$ then Eq.~(\ref{Wp}) has two real solutions:
\begin{eqnarray}
y_{1}&=&\wp(x),\label{y1}\\
y_{2}&=&\wp(x+\omega_2)\label{y2}
\end{eqnarray}
corresponding to the unbounded and bounded solutions respectively, whereas for 
$\Delta<0$ there is only one real solution given by (\ref{y1}).

\begin{figure}[t]
\centering
\includegraphics[scale=0.3]{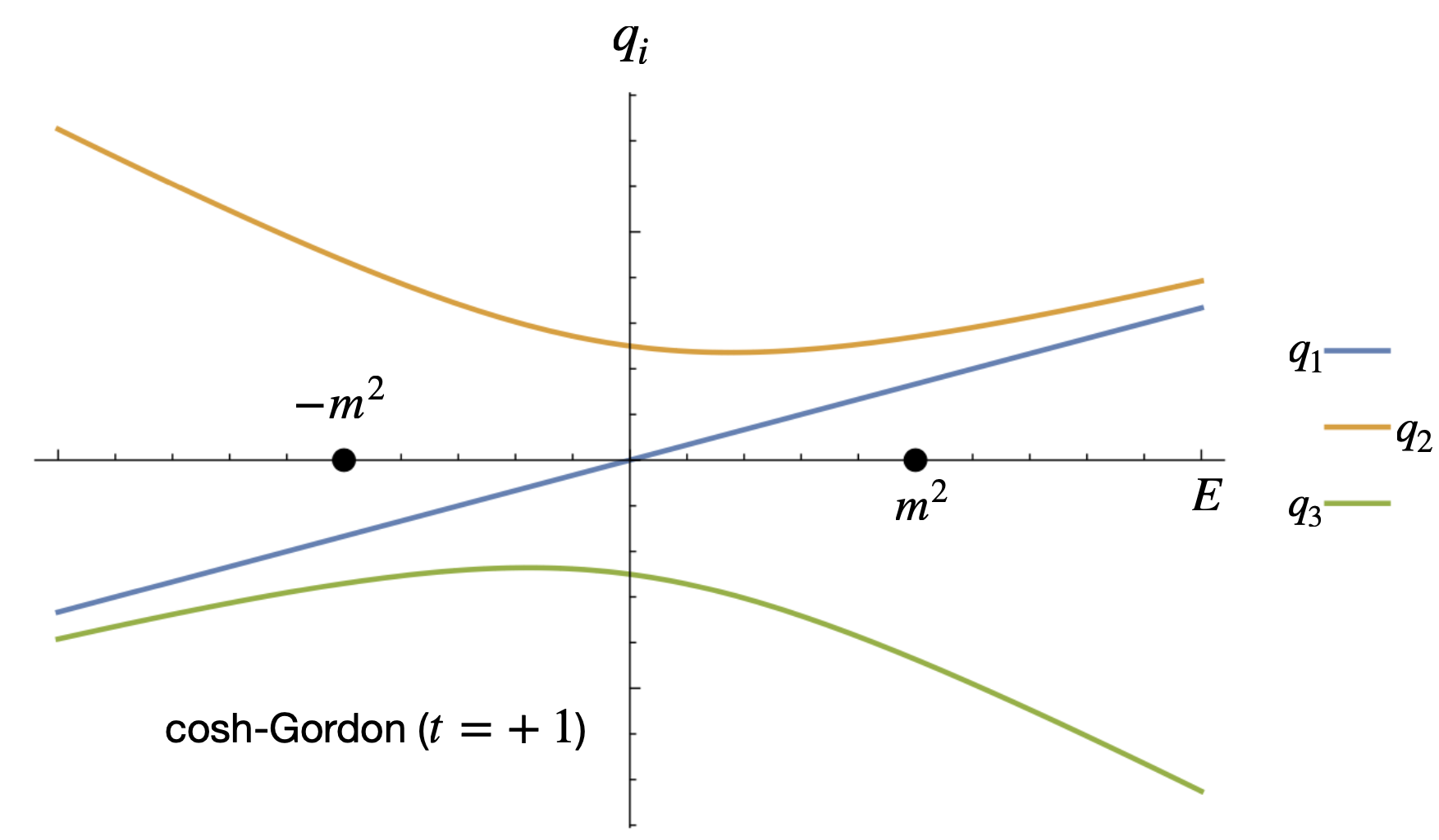}
\caption{\it The roots (\ref{roots}) for $t=+1$ as function of $E$.}
\label{fig3}
\end{figure}

\begin{figure}[t]
\centering
\includegraphics[scale=0.3]{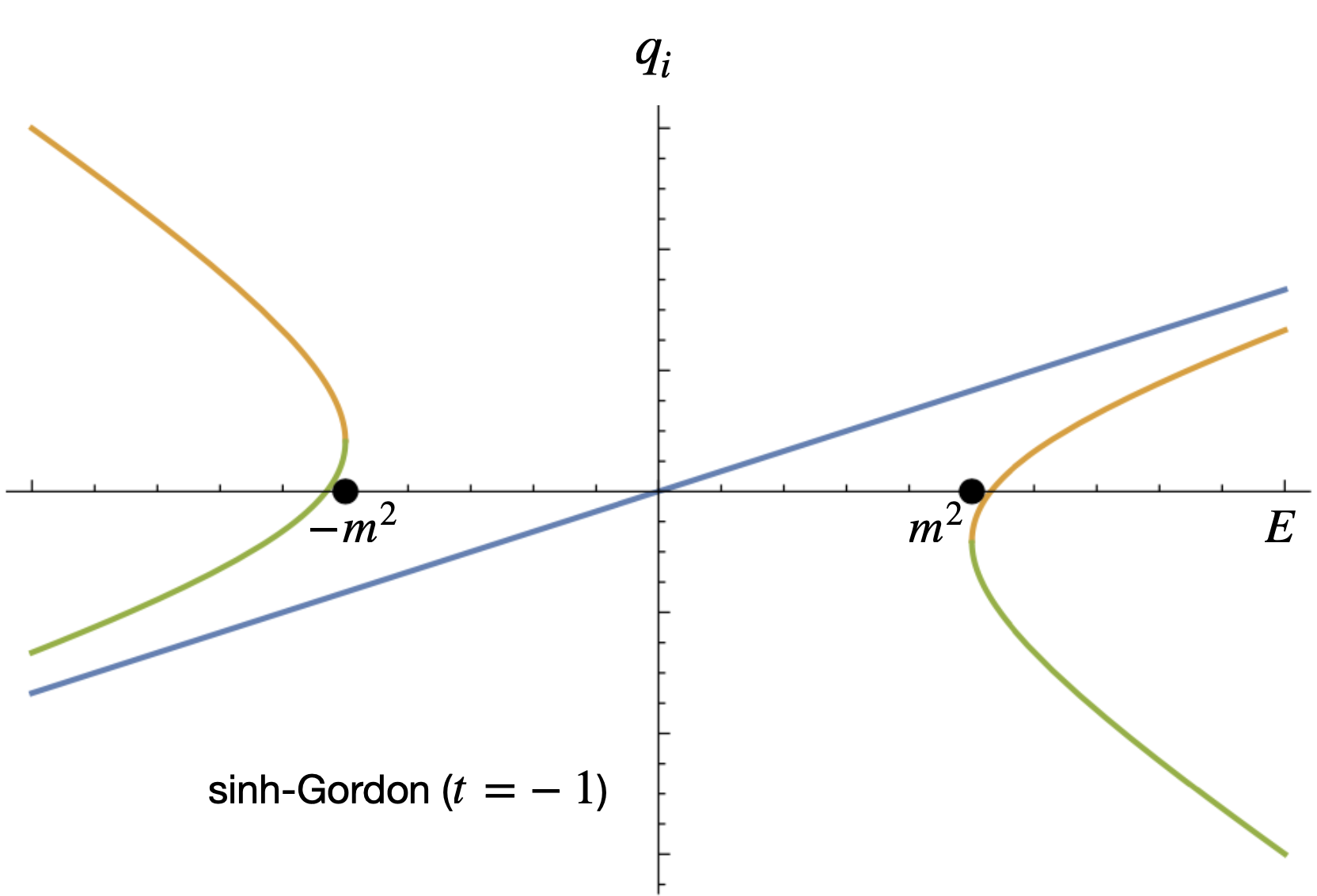}
\caption{\it The roots (\ref{roots}) for $t=-1$ as function of $E$.}
\label{fig4}
\end{figure}

At this point, we are ready to calculate:
\begin{itemize}
\item[---] the real solutions of Eq.~(\ref{Wei}); 
\item[---] the translationally invariant solutions of the sinh-/cosh-Gordon equation (\ref{OurCGeq2}); 
\item[---] analytic expressions for $\rho_{1}^{u}$ and $\rho_{2}^{u}$, $\rho_{1}^{d}$, $\rho_{2}^{d}$.
\end{itemize}
Eq.~(\ref{Wei}) is solved by
\begin{eqnarray}
y_{1}&=&\wp(x_1;g_2(E,t),g_3(E,t)),\label{y11}\\
y_{2}&=&\wp(x_1+\omega_2;g_2(E,t),g_3(E,t)).\label{y12}
\end{eqnarray}
The second solution is valid only when there are three real roots.
The coefficients $g_2$ and $g_3$ are given by (\ref{g2g3}) and related cubic polynomial is
\begin{equation}
Q(q)\;=\;4q^3-\left(\frac{1}{3}E^2+t\frac{m^4}{4}\right)q
+\frac{E}{3}\left(\frac{1}{9}E^2+t\frac{m^4}{8}\right).
\end{equation}
It is easy to obtain all three roots of $Q(q)$, they are 
\begin{equation}\label{roots}
q_1=\frac{E}{6},\qquad q_{2,3}=-\frac{E}{12}\pm\frac{1}{4}\sqrt{E^2+t m^4}.
\end{equation}
Recall, we set $e_1>e_2>e_3$ when all roots are real. If only one root is real 
this will be $e_2$, and $e_1$ will be the complex root with a positive imaginary part.
Taking into account (\ref{roots}) we have the following orderings of the roots $q_{1,2,3}(E)$:\footnote{See Fig.\ref{fig3} and  Fig.\ref{fig4}.}
\begin{enumerate}
\item[i.] if $t=+1$ then $q_2>q_1>q_3$ and $e_1=q_2$, $e_2=q_1$, $e_3=q_3$;
\item[ii.] $t=-1$, $E>m^2$ then $q_1>q_2>q_3$ and $e_1=q_1$, $e_2=q_2$, $e_3=q_3$;
\item[iii.] $t=-1$, $E<-m^2$ then $q_2>q_3>q_1$ and $e_1=q_2$, $e_2=q_3$, $e_3=q_1$;
\item[iv.] $t=-1$, $|E|<m^2$ then $e_2=q_1$ (real), $e_1=q_2$ and $e_3=q_3$ (complex).
\end{enumerate}
Going forward let us repeat it again, the unbounded solution ranges from $e_1$ to infinity 
when $\Delta>0$ and from $e_2$ to infinity when $\Delta<0$. 
The bounded solution ranges from $e_3$ to $e_2$. 
\begin{table}[h!]
	\centering
	\begin{tabular}{||c c ||} 
		\hline
		unbounded solutions & bounded solution \\ [0.5ex] 
		\hline\hline
		$y\in\left[e_1,+\infty\right)$\;{\rm for}\;$\Delta>0$\; & $y\in\left[e_3,e_2\right]$\;{\rm for}\;$\Delta>0$\;  \\
		$y\in\left[e_2,+\infty\right)$\;{\rm for}\;$\Delta<0$\; & \\ [1ex] 
		\hline
	\end{tabular}
	\caption{The ranges of real solutions of the Eq.~(\ref{Wp}), see Fig.\ref{fig1}.}
	\label{table0}
\end{table}
Then, using these informations (see Tab.\ref{table0}) and (\ref{V1}), namely,
\begin{equation}\label{v1f}
v_{1}\;=\;2y-\frac{1}{3}E\;=\;2y-\frac{2}{6}E\;=\;2(y-q_1),
\end{equation}
one gets $v_{1}$ and its ranges in all listed above cases i.--iv. (see Tab.\ref{tableV1}).
\begin{table}[h!]
	\centering
	\begin{tabular}{||c c c c||} 
		\hline
parameters	& $v_{1}$     & unbounded range                   & bounded range \\ [0.5ex] 
		\hline\hline
$t=+1$	                  & $2(y-e_2)$  & $\left[2(e_1-e_2),+\infty\right)$ & $\left[-2(e_2-e_3),0\right]$\;  \\
$t=-1$,	$E>m^2$           & $2(y-e_1)$  & $\left[0,+\infty\right)$          & $\left[-2(e_1-e_3),-2(e_1-e_2)\right]$\;  \\
$t=-1$,	$E<-m^2$          & $2(y-e_3)$  & $\left[2(e_1-e_3),+\infty\right)$ & $\left[0,2(e_2-e_3)\right]$\;  \\
$t=-1$,	$|E|<m^2$         &	$2(y-e_2)$  & $\left[0,+\infty\right)$          & ---\\ [1ex] 
		\hline
	\end{tabular}
	\caption{The ranges of $v_{1}$.}
	\label{tableV1}
\end{table}
Further, on the basis of (\ref{V1}) and (\ref{v1alfa}), one can write down for instance the explicit form of 
the sought (bounded) solutions:
\begin{eqnarray}
v_{1}(x_1;E,t)&=&2\wp(x_1+\omega_2;g_2(E,t),g_3(E,t))-\frac{E}{3},\\
\eta_{1}(x_1;E,t)&=&\log\left[\frac{2}{m^2}\left(2\wp(x_1+\omega_2;g_2(E,t),g_3(E,t))-\frac{E}{3}\right)\right].
\end{eqnarray}
Finally, using (\ref{eta}), (\ref{e1}), (\ref{v1alfa}) and (\ref{v1f}) one can calculate $\rho_{1}^{u}$, i.e., 
\begin{equation}\label{Rr11}
\rho_{1}^{u}\;=\;\sqrt{|A_1|}{\rm e}^{\eta}\;=\;\sqrt{|A_1|}{\rm e}^{\eta_1}\;=\;\frac{2\sqrt{|A_1|}}{m^2}v_{1}
\;=\;\frac{\pi}{\lambda}v_{1},
\end{equation}
and its ranges for all values of $\nu_1$ written in the Table~\ref{tableV1} (see Tab.\ref{tablerho11}).
The ranges of the remaining amplitudes $\rho_{2}^{u}$, $\rho_{1}^{d}$, $\rho_{2}^{d}$ can be calculated from (\ref{Rr11}) and
(\ref{b1}), (\ref{Asamp}): 
\begin{eqnarray}
\rho_{2}^{u}&=&\rho_{1}^{u}\;=\;\frac{\pi}{\lambda}v_1,\\
\rho_{1}^{d}&=&\rho_{2}^{d}\;=\;\frac{A_1\lambda}{\pi}\frac{1}{v_1},
\end{eqnarray}
(see Tab.\ref{tablerho11} and Tab.\ref{tablerho12}).
\begin{table}[h!]
	\centering
	\begin{tabular}{||c c c c||} 
		\hline
		parameters	& $\rho_{1}^{u}=\rho_{2}^{u}=\frac{\pi}{\lambda}v_1$ & unbounded range           & bounded range \\ [0.5ex] 
		\hline\hline
		$t=+1$	                  & $\frac{2\pi}{\lambda}(y-e_2)$  & $\left[\frac{2\pi}{\lambda}(e_1-e_2),+\infty\right)$ & $\left[-\frac{2\pi}{\lambda}(e_2-e_3),0\right]$\;  \\
		$t=-1$,	$E>m^2$           & $\frac{2\pi}{\lambda}(y-e_1)$  & $\left[0,+\infty\right)$          & $\left[-\frac{2\pi}{\lambda}(e_1-e_3),-\frac{2\pi}{\lambda}(e_1-e_2)\right]$\;  \\
		$t=-1$,	$E<-m^2$          & $\frac{2\pi}{\lambda}(y-e_3)$  & $\left[\frac{2\pi}{\lambda}(e_1-e_3),+\infty\right)$ & $\left[0,\frac{2\pi}{\lambda}(e_2-e_3)\right]$\;  \\
		$t=-1$,	$|E|<m^2$         &	$\frac{2\pi}{\lambda}(y-e_2)$  & $\left[0,+\infty\right)$          & ---\\ [1ex] 
		\hline
	\end{tabular}
	\caption{The ranges of $\rho_{1}^{u}$ and $\rho_{2}^{u}$.}
	\label{tablerho11}
\end{table}
\begin{table}[h!]
	\centering
	\begin{tabular}{||c c c c||} 
		\hline
		parameters	              &$\rho_{1}^{d}=\rho_{2}^{d}=\frac{A_1\lambda}{\pi}v_{1}^{-1}$     & unbounded range  & bounded range \\ [0.5ex] 
		\hline\hline
		$t=+1$	                  & $\frac{A_1\lambda}{2\pi}(y-e_2)^{-1}$  & $\left[\frac{A_1\lambda}{2\pi}(e_1-e_2)^{-1},0\right)$ & $\left[-\frac{A_1\lambda}{2\pi}(e_2-e_3)^{-1},{\rm sgn}(A_1\lambda)\infty\right)$\;  \\
		$t=-1$,	$E>m^2$           & $\frac{A_1\lambda}{2\pi}(y-e_1)^{-1}$  & $\left({\rm sgn}(A_1\lambda)\infty,0\right)$          & $\left[-\frac{A_1\lambda}{2\pi}(e_1-e_2)^{-1},-\frac{A_1\lambda}{2\pi}(e_1-e_3)^{-1}\right]$\;  \\
		$t=-1$,	$E<-m^2$          & $\frac{A_1\lambda}{2\pi}(y-e_3)^{-1}$  & $\left(0,\frac{A_1\lambda}{2\pi}(e_1-e_3)^{-1}\right]$ & $\left[\frac{A_1\lambda}{2\pi}(e_2-e_3)^{-1},{\rm sgn}(A_1\lambda)\infty\right)$\;  \\
		$t=-1$,	$|E|<m^2$         &	$\frac{A_1\lambda}{2\pi}(y-e_2)^{-1}$  & $\left(0,{\rm sgn}(A_1\lambda)\infty\right)$          & ---\\ [1ex] 
		\hline
	\end{tabular}
	\caption{The ranges of $\rho_{1}^{d}$ and $\rho_{2}^{d}$.}
	\label{tablerho12}
\end{table}

\subsection{Coinciding roots and hyperbolic or trigonometric solutions}
Elliptic solutions to the Eq.~(\ref{OurCGode}) with $t=-1$ have an interesting limit, 
where they are no longer given by elliptic functions. More precisely, the Weierstrass 
$\wp$-function degenerates to hyperbolic or trigonometric functions when two roots coincide. 
These solutions have been studied in \cite{Bakas2016}. We will use these results in the present subsection.
To begin with let us quote the result regarding the function $\wp(z,g_2,g_3)$ 
reported in \cite{Bakas2016}, i.e.:
\begin{enumerate}
\item
When the two larger roots coincide, $e_1=e_2=e_0$, in which case, $e_3=-2e_0$, 
the parameters $g_2$ and $g_3$ equal to $12e_{0}^{2}$ and $-8e_{0}^{3}$, respectively.
In this case, the real period $\omega_1$ tends to infinity while the imaginary period takes the value 
$\omega_2={\rm i}\pi/\sqrt{12e_0}$. Then,
\begin{equation}\label{P1}
\wp\!\left(z,12e_{0}^{2},-8e_{0}^{3}\right)\;=\;e_0\left(1+\frac{3}{\sinh^2\left(\sqrt{3e_0}z\right)}\right).
\end{equation}
\item 
When the two smaller roots coincide, $e_2=e_3=-e_0$, in which case $e_1=2e_0$, the parameters  
$g_2$ and $g_3$ read as follows $g_2=12e_{0}^{2}$, $g_3=8e_{0}^{3}$. This time the imaginary period
$\omega_2$ diverges and the real period is $\omega_1=\pi/\sqrt{12e_0}$.
Then,
\begin{equation}\label{P2}
\wp\!\left(z,12e_{0}^{2},8e_{0}^{3}\right)\;=\;e_0\left(-1+\frac{3}{\sin^2\left(\sqrt{3e_0}z\right)}\right).
\end{equation}
\end{enumerate}
Looking at Fig.\ref{fig3} and Fig.\ref{fig4}, and the orderings of the roots (points i.-iv.), 
one may see two possibilities for degeneracy, i.e., in cases with $t=-1$, namely, for $E=-m^2$ and $E=m^2$.
Below, we calculate solutions corresponding to these situations.

\subsubsection*{${\bf E=-m^2}$} 
In this case the roots are $e_1=e_2=e_0=\frac{1}{12}m^2$ and $e_3=-\frac{1}{6}m^2$.
The real half-period $\omega_1$ diverges and the imaginary half-period becomes $\omega_2={\rm i}\pi/m$.
Applying (\ref{P1}) to the solution (\ref{y11}) one gets
\begin{itemize}
\item the function $v_1$,
\begin{eqnarray}
v_{1}(x_1)&=&2\wp(x_1)-\frac{E}{3}\;=\;2\left[\frac{m^2}{12}\left(1+\frac{3}{\sinh^2\left(\frac{1}{2}mx_1\right)}\right)\right]
+\frac{m^2}{3}\nonumber
\\
&=&\frac{m^2}{2}{\coth}^2\!\left(\frac{1}{2}mx_1\right) \;=\;\frac{\lambda\sqrt{|A_1|}}{\pi}
{\coth}^2\!\left(\frac{1}{2}\sqrt{\frac{2\lambda\sqrt{|A_1|}}{\pi}}x_1\right);
\end{eqnarray}
\item the field $\eta_1$,
\begin{eqnarray}\label{kink}
\eta_{1}(x_1)&=&\log\frac{2}{m^2}v_{1}(x_1)\;=\;2\log\coth\!\left(\frac{1}{2}mx_1\right)\nonumber
\\
&=&
2\log\coth\!\left(\frac{1}{2}\sqrt{\frac{2\lambda\sqrt{|A_1|}}{\pi}}x_1\right);
\end{eqnarray}
\item the densities:
\begin{eqnarray}
\rho_{1}^{u}(x_1)&=&\rho_{2}^{u}(x_1)\;=\;\frac{\pi}{\lambda}v_1(x_1)\nonumber
\\
&=&\sqrt{|A_1|}{\coth}^2\!\left(\frac{1}{2}\sqrt{\frac{2\lambda\sqrt{|A_1|}}{\pi}}x_1\right),
\\
\rho_{1}^{d}(x_1)&=&\rho_{2}^{d}(x_1)\;=\;\frac{A_1\lambda}{\pi}\frac{1}{v_1(x_1)}\nonumber
\\
&=&\frac{A_1}{\sqrt{|A_1|}}{\coth}^{-2}\!\left(\frac{1}{2}\sqrt{\frac{2\lambda\sqrt{|A_1|}}{\pi}}x_1\right).
\end{eqnarray}
\end{itemize}
Applying (\ref{P1}) to the solution (\ref{y12}) one obtains
\begin{itemize}
\item the function $v_1$,
\begin{eqnarray}
v_{1}(x_1)&=&2\wp(x_1+\omega_2)-\frac{E}{3}
\;=\;2\left[\frac{m^2}{12}\left(1+\frac{3}{\sinh^2\left(\frac{1}{2}mx_1+\frac{1}{2}m\omega_2\right)}\right)\right]
+\frac{m^2}{3}\nonumber
\\
&=&\frac{m^2}{2}{\tanh}^2\!\left(\frac{1}{2}mx_1\right) \;=\;\frac{\lambda\sqrt{|A_1|}}{\pi}
{\tanh}^2\!\left(\frac{1}{2}\sqrt{\frac{2\lambda\sqrt{|A_1|}}{\pi}}x_1\right);
\end{eqnarray}
	\item the field $\eta_1$,
	\begin{eqnarray}\label{antikink}
		\eta_{1}(x_1)&=&\log\frac{2}{m^2}v_{1}(x_1)\;=\;2\log\tanh\!\left(\frac{1}{2}mx_1\right)\nonumber
		\\
		&=&
		-2\log\coth\!\left(\frac{1}{2}\sqrt{\frac{2\lambda\sqrt{|A_1|}}{\pi}}x_1\right);
	\end{eqnarray}
	\item the densities:
	\begin{eqnarray}
		\rho_{1}^{u}(x_1)&=&\rho_{2}^{u}(x_1)\;=\;\frac{\pi}{\lambda}v_1(x_1)\nonumber
		\\
		&=&\sqrt{|A_1|}{\tanh}^2\!\left(\frac{1}{2}\sqrt{\frac{2\lambda\sqrt{|A_1|}}{\pi}}x_1\right),
		\\
		\rho_{1}^{d}(x_1)&=&\rho_{2}^{d}(x_1)\;=\;\frac{A_1\lambda}{\pi}\frac{1}{v_1(x_1)}\nonumber
		\\
		&=&\frac{A_1}{\sqrt{|A_1|}}{\tanh}^{-2}\!\left(\frac{1}{2}\sqrt{\frac{2\lambda\sqrt{|A_1|}}{\pi}}x_1\right).
	\end{eqnarray}
\end{itemize}

\subsubsection*{${\bf E=m^2}$} 
In such a situation the roots are $e_1=\frac{1}{6}m^2$ and $e_2=e_3=-e_0=-\frac{1}{12}m^2$. The imaginary 
half-period $\omega_2$ diverges and the real half-period is $\omega_1=\pi/m$.
This time, applying the formula (\ref{P2}) to Eq.~(\ref{y11}) one gets
\begin{itemize}
\item the function $v_1$,
\begin{eqnarray}
v_{1}(x_1)&=&2\wp(x_1)-\frac{E}{3}\;=\;2\left[\frac{m^2}{12}\left(-1+\frac{3}{\sin^2\left(\frac{1}{2}mx_1\right)}\right)\right]
-\frac{m^2}{3}\nonumber
\\
&=&\frac{m^2}{2}{\cot}^2\!\left(\frac{1}{2}mx_1\right) \;=\;\frac{\lambda\sqrt{|A_1|}}{\pi}
{\cot}^2\!\left(\frac{1}{2}\sqrt{\frac{2\lambda\sqrt{|A_1|}}{\pi}}x_1\right);
\end{eqnarray}
	\item the field $\eta_1$,
	\begin{eqnarray}
		\eta_{1}(x_1)&=&\log\frac{2}{m^2}v_{1}(x_1)\;=\;2\log\cot\!\left(\frac{1}{2}mx_1\right)\nonumber
		\\
		&=&
		2\log\cot\!\left(\frac{1}{2}\sqrt{\frac{2\lambda\sqrt{|A_1|}}{\pi}}x_1\right);
	\end{eqnarray}
	\item the densities:
	\begin{eqnarray}
		\rho_{1}^{u}(x_1)&=&\rho_{2}^{u}(x_1)\;=\;\frac{\pi}{\lambda}v_1(x_1)\nonumber
		\\
		&=&\sqrt{|A_1|}{\cot}^2\!\left(\frac{1}{2}\sqrt{\frac{2\lambda\sqrt{|A_1|}}{\pi}}x_1\right),
		\\
		\rho_{1}^{d}(x_1)&=&\rho_{2}^{d}(x_1)\;=\;\frac{A_1\lambda}{\pi}\frac{1}{v_1(x_1)}\nonumber
		\\
		&=&\frac{A_1}{\sqrt{|A_1|}}{\cot}^{-2}\!\left(\frac{1}{2}\sqrt{\frac{2\lambda\sqrt{|A_1|}}{\pi}}x_1\right).
	\end{eqnarray}
\end{itemize}

As a final remark in this subsection let us note that among the solutions listed above, one can find soliton-like structures such as (\ref{kink}) and (\ref{antikink}).
They can be seen as ``superpositions'' of soliton-like profiles.\footnote{The function $-\log(\cosh(\cdot))$ describes, for instance,
the profile of a dilaton field in a two-dimensional dilaton gravity, see~\cite{LA,BS}.}

\section{Concluding remarks}
\label{Sect4}
In \cite{F1} and \cite{FPPZ(2019)NPB} a connection has been established between polymers and the statistical mechanics of non-relativistic anyon particles. While this connection is intriguing, many of its aspects remained unclear. First of all, the equations of motion of anyon particles admit self-dual static solutions that minimize their energy. So far, it was unclear what is the meaning of self-duality from the polymer point of view. Second, explicit self-dual configurations minimizing the energy (\ref{energy1}) were missing. That energy resembles that of the Abelian Higgs model in the limit in which the density of half of the fields is constant, see Eq.~(\ref{energy2}), but is more complicated in the general case.
Within this work, we have considered the splitting established in \cite{F1}
of the polymer action of a 4-plat composed by two linked rings
into a self-dual part $I_{\sf sd}$ and a non-self-dual contribution  $I_{\sf C}$.
It has been noticed that this splitting corresponds to two different kinds of the interactions of entropic origin that are present in a polymer system that is subjected to topological constraints.
The term $I_{\sf C}$ is responsible for Coulomb-like interactions.
The potentials in $I_{\sf C}$ describe local interactions that can be both attractive or repulsive.
The repulsive component is necessary to prevent that the polymer lines
cross themselves, as this would break the topology of the link.
On the other side, the more the two rings are linked together, the more their centers of mass are getting closer. This phenomenon is well known \cite{Quake,FFIL1999} and explains why topological constraints are responsible also for attractive forces between the monomers.
Of course, topological properties are
global in nature, so that local interactions are not enough to preserve them.
The self-dual term $I_{\sf sd}$ accounts for the long range interactions
that are necessary to globally enforce the topological relations.
Indeed, the Chern-Simons vector potentials ${\boldsymbol B}_a$, $a=1,2$, 
that mediate such interactions, appear only inside $I_{\sf sd}$.

The self-dual regime occurs
when the rings are homopolimeric, see Eq.~(\ref{homcond}).
We have shown using a scaling argument that
in the limit $\tau\to\infty$ 
static self-dual solutions
of the model described by the action (\ref{Sanyon}) or equivalently (\ref{action3})
make sense also in the polymer case and not only in its anyon particles interpretation.
In this limit the height of the 4-plat becomes infinite and the monomer density becomes independent of $z$.

Next, the energy landscape of the 4-plat has been explored. It turns out that it is quite complex as it is characterized by at least two points of minimum
in the simplified situation in which the monomer density of the legs of the rings point downwards (see Fig.\ref{fig4plat}) 
is assumed to be constant.

As an application of the established polymer-anyon model, classes of explicit  solutions of its self-dual equations have been derived.
First, we show that  the densities of monomers $\left|\psi^u_a\right|^2$ and $|\psi^d_a|^2$ for the components $\Gamma_a^{u,d}$ of the loops $\Gamma_a$, $a=1,2$, may be described by  holomorphic functions provided  $|\psi^u_a|^2=|\psi^d_a|^2$.
As a further step, we prove that, under some assumptions, the self-dual equations (\ref{sdcond}) may be identified with the 
sinh-Gordon or cosh-Gordon equations for the monomer densities (\ref{SGE})-(\ref{CGE}). To some extent, it is possible to conclude that 
Eq.~(\ref{SGE}) is the analog of the Gouy-Chapman equation \cite{Chen,Gouy,Chapman}, but instead of the spatial distribution of the electric potential of charged particles, our equation describes the spatial distribution of the fictitious magnetic fields associated with the presence of the topological constraints.
In the limit in which the two spatial dimensions are large in comparison with the third one, we provide exact formulas fo the conformations of the monomer densities by using the elliptic, hyperbolic and trigonometric solutions of the sinh-Gordon and cosh-Gordon equations which have been used for instance in the  construction of classical string solutions in AdS3 and dS3 \cite{Bakas2016}.

\section*{Acknowledgments}
The research presented here has been supported
by the Polish National Science Centre under grant no. 2020/37/B/ST3/01471. This work
results within the collaboration of the COST Action CA17139 (EUTOPIA).

\end{document}